\documentclass[aps,pre,showpacs,floats,floats,twocolumn,superscriptaddress,floatfix,english]{revtex4-1}

\usepackage[T1]{fontenc}
\usepackage[latin9]{inputenc}
\usepackage{amsmath}
\usepackage{amssymb}
\usepackage{graphicx}

\usepackage{graphicx}
\usepackage{epstopdf, epsfig}
\usepackage{dutchcal}
\usepackage{array}
\usepackage{multirow}
\usepackage{setspace}
\usepackage{subfigure}
\usepackage{caption}
\usepackage{color}
\usepackage{epstopdf}
\usepackage[normalem]{ulem}
\usepackage{float}
\usepackage{chemformula}
\usepackage{siunitx}

\DeclareMathAlphabet\mathzapf{T1}{pzc}{mb}{it}
\DeclareMathAlphabet\mathrsfso{U}{rsfso}{m}{n}

\newcommand{\bb}{\boldsymbol}
\newcommand{\del}{\partial}
\DeclareMathOperator{\Log}{Log} 
\DeclareMathOperator{\li}{li}

\newcommand{\cauchy}{\mathzapf{P}} 
\newcommand{\dair}{\rho_{\rm{a}}} 
\newcommand{\dwat}{\rho_{\rm{w}}} 
\newcommand{\kc}{k_{\rm{cap}}} 
\newcommand{\lc}{l_{\rm{cap}}} 
\newcommand{\cm}{c_{\rm{min}}} 
\newcommand{\lin}{\mathrsfso{L}} 
\newcommand{\K}{\mathzapf{K}} 
\newcommand{\V}{\mathzapf{V}} 
\newcommand{\E}{\mathzapf{E}} 
\newcommand{\zex}{\mathcal{z}_{\star}} 
\newcommand{\zc}{z_{\rm{c}}} 
\newcommand{\crit}{\chi_{\rm{c}}} 

\newcommand{\z}{\mathcal{z}} 
\newcommand{\kd}{\mathcal{k}} 
\newcommand{\U}{\mathcal{U}} 
\newcommand{\Upc} {\U'_{\rm{c}}}  
\newcommand{\Uppc} {\U''_{\rm{c}}}  
\newcommand{\C}{\mathcal{C}} 
\newcommand{\kcd}{\mathcal{k}_{\rm{cap}}} 
\newcommand{\cmd}{\mathcal{C}_{\rm{min}}} 
\newcommand{\zcd}{\mathcal{z}_{\rm{c}}} 
\newcommand{\zs}{\mathcal{z}_{\rm{s}}} 
\newcommand{\out}{\chi_{\rm{out}} } 
\newcommand{\solinf}{\chi_{\infty}} 
\newcommand{\X}{\mathcal{X}} 
\newcommand{\Z}{\mathzapf{Z}} 
\newcommand{\Xout}{\mathcal{X}_{\rm{out}}} 
\newcommand{\Zs}{\mathzapf{Z}_{\rm{s}}} 
\newcommand{\Xinf}{\mathcal{X}_{\infty}} 
\newcommand{\Xu}{\mathcal{X}_{\rm{unif}}} 
\newcommand{\Long}{\hat{\epsilon}} 
\newcommand{\flex}{\check{\epsilon}} 
\newcommand{\kmax}{\kd_{\star}} 
\newcommand{\xmax}{\mathzapf{x}_{\star}} 
\newcommand{\gm}{\gamma_{\rm{max}}}  
\newcommand{\pp}{\mathcal{p}} 

\newcommand{\psir}{\hat{\psi}_{\rm{r}}}  
\newcommand{\psii}{\hat{\psi}_{\rm{i}}} 
\newcommand{\x}{\tilde{x}} 
\newcommand{\xe}{\tilde{x}_{\star}} 
\newcommand{\ze}{z_{\star}} 
\newcommand{\Ze}{Z_{\star}} 
\newcommand{\xeo}{\tilde{x}_{1\star}} 
\newcommand{\xet}{\tilde{x}_{2\star}} 
\newcommand{\xej}{\tilde{x}_{j\star}} 

\newcommand{\nn}{\nonumber}
\newcommand{\be}{\begin{equation}}
\newcommand{\ee}{\end{equation}}
\newcommand{\bc}{\begin{cases}}
\newcommand{\ec}{\end{cases}}

\newcommand{\nordita}{Nordita, KTH Royal Institute of Technology and Stockholm University, Stockholm 10691, Sweden}

\begin{document}

\title{Asymptotic interpretation of the Miles mechanism of wind-wave instability}

\author{A. F. Bonfils}
\affiliation{\nordita}
\author{Dhrubaditya Mitra}
\affiliation{\nordita}
\author{W. Moon}
\affiliation{\nordita}
\affiliation{Department of Mathematics, Stockholm University 106 61, Sweden}
\author{J. S. Wettlaufer}
\affiliation{\nordita}
\affiliation{Yale University, New Haven, Connecticut 06520-8109, USA}

\makeatother

\begin{abstract}
When wind blows over water, ripples are generated on the water surface. These ripples can be regarded as perturbations of the wind field, which is modelled as a parallel inviscid flow. For a given wavenumber $k$, the perturbed streamfunction of the wind field and the complex phase speed are the eigenfunction and the eigenvalue of the so-called Rayleigh equation in a semi-infinite domain. Because of the small air--water density ratio, $\dair/\dwat\equiv\epsilon\ll1$, the wind and the ripples are weakly coupled, and the eigenvalue problem can be solved perturbatively. At the leading order, the eigenvalue is equal to the phase speed $c_0$ of surface waves. At order $\epsilon$, the eigenvalue has a finite imaginary part, which implies growth. \citet{miles57} showed that the growth rate is proportional to the square modulus of the leading-order eigenfunction evaluated at the so-called critical level $z=z_c$, where the wind speed is equal to $c_0$ and the waves extract energy from the wind. Here, we construct uniform asymptotic approximations of the leading-order eigenfunction for long waves, which we use to calculate the growth rate as a function of $k$. In the strong wind limit, we find that the fastest growing wave is such that the aerodynamic pressure is in phase with the wave slope. The results are confirmed numerically. 
\end{abstract}
\maketitle
\begin{figure}
  \centerline{\includegraphics[scale=0.175]{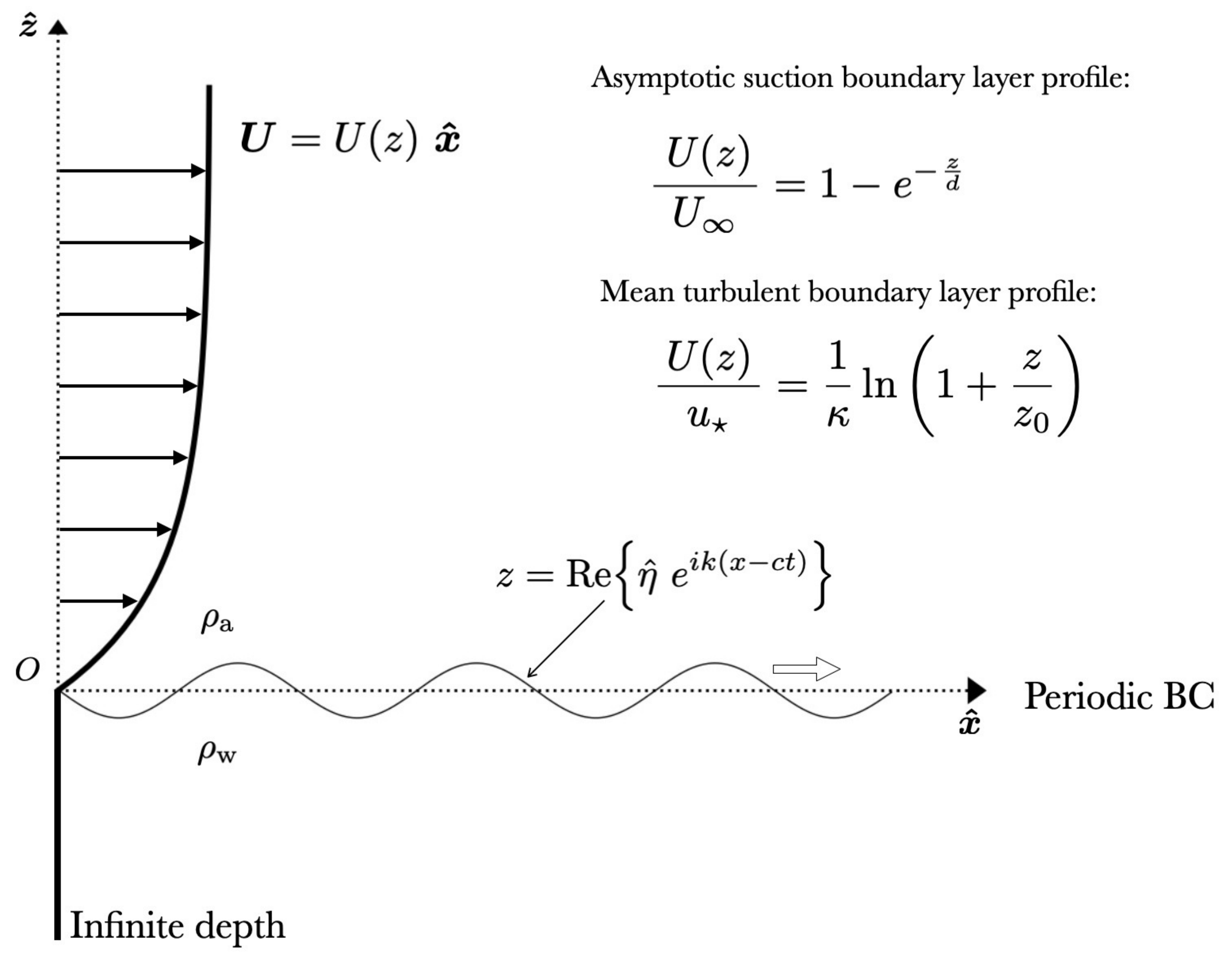}}
  \caption{Schematic of the mean wind field and a normal mode of the air--water interface. In the exponential profile, $U_{\infty}$ is the far field wind velocity and $d$ the thickness of the air boundary layer. In the logarithmic profile, $\kappa=0.4$ is the von K\'arm\'an constant, $u_*$ the friction velocity of the wind field and $z_0$ a roughness length accounting for the presence of ripples on the water surface.}
\label{schema}
\end{figure}
\section{Introduction}
The generation and the growth of water waves by wind is an old problem in geophysical fluid dynamics, with a wide range of applications, and challenges that have occupied the community for at least 150 years \citep{helmholtz1868, kelvin1871}. \citet{jeffreys25} suggested that wind waves grow because the pressure on the windward face of a crest is greater than the pressure on the leeward face of that crest, an ansatz he called the `sheltering hypothesis'. The modern foundations of a theory were laid down by \citet{phillips57} and \citet{miles57}, summaries of which can be found in the books of \citet{phillipsbook} and \citet{janssenbook}. 

We consider a layer of water of infinite depth over which a turbulent wind blows (Fig.~\ref{schema}). The air pressure fluctuations generate ripples on the water surface \citep{phillips57}. However, because the generation time scale is much smaller than the ripple period, we average the turbulent fluctuations over the longest period and model the mean wind field as a parallel inviscid steady flow, $\bb{U} = U(z)\ \bb{\hat{x}}$, where $U$ is a continuous and monotonic function of the vertical coordinate, $z$, and $\bb{\hat{x}}$ is a horizontal unit vector. Following \citet{miles57}, we study the linear stability of the wind field under perturbations induced by the tiny waves -- which we call ripples -- generated by turbulent fluctuations on the water surface, including gravity, $g$, and surface tension, $\sigma$. The shear is efficiently dissipated in the water, so that $U(z\le 0)=0$. We restrict our analysis to two-dimensional incompressible perturbations, assuming that Squire's theorem holds (we check it a posteriori in Appendix \ref{Appsquire}). The amplitude of a wave-induced perturbation as a function of the vertical variable, $z$, is determined by the Rayleigh equation, which expresses the conservation of vorticity along the streamlines \citep{drazin-reid}. 

Key quantities to determine are the Fourier components of the aerodynamic pressure, which \citet{miles57} wrote as
\be
\hat{p}_0(0^+)\equiv \dair V^2 (\alpha + i \beta) k \hat{\eta}_0,\quad\text{with}\quad\alpha,\beta =O(1), \label{alphabeta}
\ee
where $\dair$ is the density of air, $V$ is a characteristic wind speed and $\hat{\eta}_0$ is the amplitude of a Fourier mode --  characterized by the wavenumber $k$ -- of the displacement of the water surface; the subscript $0$ indicates that these are leading-order quantities in the expansion in powers of the air/water density ratio, 
\be
{\frac{\dair}{\dwat}\equiv\epsilon\ll1.}
\ee
We emphasize that equation (\ref{alphabeta}) is a generalization of the Jeffreys sheltering hypothesis, which states that the aerodynamic pressure is in phase with the wave slope, and thus $\alpha =0$. The calculation of $\alpha$ and $\beta$ involves the solution of the Rayleigh equation, which exhibits singular behaviour at the critical level $z=\zc$, where the wind velocity, $U(z)$, equals the phase speed of water waves. The problem has been studied extensively over the last 60 years with a focus on $\beta$, because it is proportional to the growth rate of the wave. \citet{conte-miles}, \citet{hughes-reid} and \citet{beji-nadaoka} solved the Rayleigh equation numerically for various wind profiles, but an exact analytical solution exists only for an exponential profile -- a crude approximation of the mean turbulent wind. Moreover, it involves a hypergeometric function from which it is difficult to extract the maximum growth rate \citep{young-wolfe}. \citet{miles93} revisited his original work using the logarithmic wind profile and including the effects of turbulence. Using a variational method, confirmed by matched asymptotic expansions, he found an approximate formula for $\beta$ and fitted a subset of the experimental growth rates collated by \citet{plant82}. 
The coefficient $\alpha$ has generally been assumed to be negative and is neglected, evidently computed only by \citet{conte-miles} and \citet{milespart2}. However, we have been unable to demonstrate that $\alpha<0$ and our analysis shows that such an assumption is false.
 
\citet{lighthill62} showed that the energy transfer from the wind to the waves occurs at the critical level, which is the height at which the wind speed equals the phase speed of water waves, observed for example by \citet{hristov-nature} in the range $16<c / u_{\star}<40$. Furthermore, in order to approximate the growth rate of the Miles instability, \citet{carpenter-et-al17} modelled the air--water interface and the critical level as interacting vortex sheets. However, their minimal model does not yield the dependence of the maximum growth rate on the physical parameters.
 
In \S \ref{model}, we describe the normal modes of the air--water interface in the presence of a wind field. We then recover the results of Miles' theory perturbatively using a small air--water density ratio; $\epsilon\ll1$.  
In \S \ref{longwaves}, we use asymptotic methods to solve the Rayleigh equation for waves whose wavelength is much larger than the characteristic length scale of a given wind profile. We note that, in an Appendix to \citet{morland-saffman}, Miles used such a long wave approximation to simplify the exact solution for an exponential wind profile, but because we work directly with the Rayleigh equation, our approach is more general. We check the accuracy of our asymptotic expressions numerically, using a variant of the method proposed by \citet{hughes-reid}. In \S \ref{appli}, we obtain explicit expressions for $\alpha$ and $\beta$, and show that $\alpha$ can be non-negative. Next, we determine the growth rate of the Miles instability, and fit the entire range of the data compiled by \citet{plant82} using the logarithmic profile. In \S \ref{strongwind}, we study the strong wind limit introduced by \citet{young-wolfe}. We find that the fastest growing wave is characterized by $\alpha=0$ and is therefore accompanied by an aerodynamic pressure that is proportional to the wave slope, consistent with the Jeffreys sheltering hypothesis. We note that this result also holds approximately for moderate wind. We conclude in \S \ref{ccl}. 

\section{Wind-wave model}  \label{model}
Ripples on the water surface create small perturbations in the wind field. The perturbed velocity field is $\bb{U} + \bb{u}$, with $\bb{u}=u(x,z,t)\ \bb{\hat{x}} + w(x,z,t)\ \bb{\hat{z}}$, where $t$ is time. The incompressibility condition, $\bb{\nabla}\cdot\bb{u}=0$, allows us to introduce the streamfunction, $\psi(x,z,t)$, such that $u =\del_z\psi$ and $w =-\del_x\psi$.
\subsection{Normal modes}
We consider a surface displacement field of the form $\eta(x,t) = \Re\big\{\hat{\eta}\ e^{ik(x-c t)}\big\}$, where $c$ is a complex phase speed to be determined. The $x$-average over a wavelength, $2\pi/k$, is denoted by an overbar. Thus, since $\overline{\eta(x,t)}=0$, the unperturbed water surface, $z=0$, corresponds to the mean water level. Following \citet{young-wolfe}, we define the wave energy, $\E\equiv\K+\V$, as the sum of the mean kinetic energy per unit area, $\K$, and the mean potential energy per unit area, $\V$, which are given by
\be
\K(t)\equiv\ \int_{-\infty}^{0^-}  dz\ \frac{\dwat \overline{|\bb{u}|^2}}{2} + \int_{0^+}^{+\infty}  dz\ \frac{\dair \overline{|\bb{u}|^2}}{2}
\ee
and 
\be
\V(t)\equiv\frac{1}{2}\overline{\Big\{ (\dwat -\dair)g\ \eta^2 + \sigma\ (\del_x\eta)^2\Big\}}, 
\ee
where $\dwat$ is the density of water. \textcolor{black}{The rate of change of wave energy is
\be
\frac{d \E}{dt} = \int_{0^+}^{+\infty} dz\ \tau(z,t)  U'(z), 
\label{NRJconserv}
\ee
and $\tau\equiv-\dair\ \overline{uw}$ is the wave-induced Reynolds stress.}

Following the canonical procedure \citep[e.g.,][]{drazin-reid}, we write the streamfunction in terms of normal modes as $\psi(x,z,t) = \Re\big\{\hat{\psi}(z)\ e^{ik(x-c t)}\big\}$. This leads to the Rayleigh equation
\be
\lin\hat{\psi} =0,\quad\text{with}\quad \lin(z,c)=\big[U(z)-c\big]\bigg[\frac{d^2}{dz^2}- k^2\bigg] - U''(z),\label{Rayleigh}
\ee
where the prime denotes differentiation with respect to $z$. The solution of equation (\ref{Rayleigh}) in the water, where there is no shear, is $\hat{\psi}(z\le 0) = \hat{\psi}(z=0)\ e^{kz} $, which we use to derive the boundary condition at $z=0^+$ and obtain \citep{morland-saffman}
\be
\Big(kc^2-g - \frac{\sigma}{\dwat} k^2\Big)\hat{\psi}(0)=\epsilon\Big\{ c^2 \hat{\psi}' + (cU'-g)\hat{\psi}\Big\}\Big|_{0^+}.\label{BC}
\ee
\textcolor{black}{\subsection{Perturbative resolution of the eigenvalue problem}} \label{perturbationtheory}
Following \citet{janssenbook} and \citet{young-wolfe}, we expand the eigenvalue and the eigenfunction in the air in a power series in $\epsilon\ll 1$ as
\refstepcounter{equation}
$$
c=c_0+\epsilon\ c_1+ ...\qquad\text{and}\qquad\hat{\psi}^{\rm{a}}=\hat{\psi}_0+\epsilon\ \hat{\psi}_1+ ...\ , \label{expansions}
  \eqno{(\theequation{\mathit{a},\mathit{b}})}
$$
where `$\rm{a}$' denotes `air'. Similarly, the amplitude of the surface displacement, $\hat{\eta}$, and the amplitude of the perturbation pressure in the air, $\hat{p}^{\rm{a}}=\hat{p}^{\rm{a}}(z)$, are 
\refstepcounter{equation}
$$
\hat{\eta}=\hat{\eta}_0+\epsilon\ \hat{\eta}_1+ ...\qquad\text{and}\qquad\hat{p}^{\rm{a}}=\hat{p}_0+\epsilon\ \hat{p}_1+ ...\ .
  \eqno{(\theequation{\mathit{a},\mathit{b}})}
$$
To leading order the ripples are not affected by the wind, but they induce a neutral perturbation on the air flow, determined by
\be
\lin(z,c_0)\ \hat{\psi}_0(z) =0,\qquad z\ge 0. \label{L_0}
\ee
The leading-order eigenvalue, $c_0$, is by definition the phase speed of water waves. Imposition of the boundary condition (\ref{BC}) yields the dispersion relation
\be
c_0(k)=\frac{\cm}{\sqrt{2}}\sqrt{\frac{\kc}{k}+\frac{k}{\kc}}\ , \label{surf1}
\ee
where  
\refstepcounter{equation}
$$
\cm\equiv \bigg[\frac{4\sigma g}{\dwat}\bigg]^{\frac{1}{4}}\qquad\text{and}\qquad \kc\equiv\sqrt{\frac{\dwat g}{\sigma}}.
  \eqno{(\theequation{\mathit{a},\mathit{b}})}\label{surf2}
$$
The minimum phase speed, $\cm$, arises from the competition between surface tension and gravitational forces and occurs when $k=\kc$; the capillary wavenumber.

Following \citet{phillipsbook}, the leading-order amplitude of the aerodynamic pressure  (cf. Eq. \ref{alphabeta}) is
\be
\hat{p}_0(0^+)= \dwat c_0^2 (\mu+ i\gamma) k \hat{\eta}_0,\quad\text{with}\quad\mu, \gamma =O(\epsilon). \label{phillips}
\ee
The phase difference between the aerodynamic pressure and the wave slope is proportional to $\mu$, which can be considered as the deviation from Jeffreys' sheltering hypothesis. Because
\be
\hat{p}_0 = \dair\ \mathrsfso{W}(\hat{\psi}_0, U-c_0), 
\ee
where $\mathrsfso{W}$ is the Wronskian, the eigenvalue at the next order -- determined by the boundary condition (\ref{BC}) -- can be written as
\be
\epsilon\ c_1 = \frac{c_0}{2}\Bigg( \mu +i \gamma -\frac{\epsilon}{1+\big[\frac{k}{\kc}\big]^2}\Bigg). \label{c1}
\ee
Hence, $\mu$ is twice the wind-dependent relative change of the phase speed of water waves due to the coupling with the air, and $\gamma$ is the energy growth rate, normalized by the angular frequency of water waves. The last term in equation (\ref{c1}), which did not appear in \citet{miles57}, is the difference between the phase speed of interfacial waves and the phase speed of surface waves. Moreover, if we expand the wave energy as
\be
\E = \E_0+\epsilon\ \E_1 + ...\ ,
\ee
we find
\be
\gamma= \frac{1}{k c_0} \frac{\epsilon}{\E_0}\frac{d\E_1}{dt}\bigg|_{t=0}. 
\label{gamma1}
\ee
Now, comparing equations (\ref{gamma1}) with (\ref{NRJconserv}), we retrieve the result of \citet{janssenbook};
\be
\gamma=\frac{\hat{\tau}_0(z=0^+)}{k\E_0},\quad\text{where}\quad \hat{\tau}_0(z) = - \dair \ \frac{k}{2}\ \Im\big\{ \hat{\psi}_0(z) \hat{\psi}'^{*}_0(z)\big\} \label{tau0}
\ee
is the leading-order amplitude of $\tau(z,t)$, in which the star denotes complex conjugation, and $\E_0$ is the energy (per unit area) of water waves. This demonstrates that water waves extract energy from the wind through the work of the wave-induced Reynolds stress. 
\subsection{Boundary-value problems}
\begin{table}
\begin{center}
\resizebox{\columnwidth}{!}{\begin{tabular} {c c c c}
\hline
\raisebox{1.5ex}{\textcolor{white}{l}}&Gravity waves&Capillary waves& Capillary-gravity waves\\[0.5ex]
\hline
\raisebox{2ex}{\textcolor{white}{l}}Control parameters&$Fr\equiv \frac{V}{\sqrt{gL}}$ &$We\equiv \frac{\dwat V^2 L}{\sigma}$&$\cmd\equiv \frac{\cm}{V}$ and $\kcd\equiv \kc L$    \\ [1ex]
\hline
\raisebox{2ex}{\textcolor{white}{l}}$\C(\kd)$&$\frac{1}{Fr\sqrt{\kd}}$ &$ \sqrt{\frac{\kd}{We}}$& $\frac{\cmd}{\sqrt{2}}\sqrt{\frac{\kcd}{\kd}+\frac{\kd}{\kcd}}$   \\ [1ex]
\hline
\raisebox{1.5ex}{\textcolor{white}{l}}$m$& $\frac{1}{Fr^2}$ &$ \frac{1}{We}$& $\frac{\cmd^2}{2}$   \\ [0.5ex]
\hline  
\raisebox{1.5ex}{\textcolor{white}{l}}$q$&$\frac{2}{3}$ &$2$& $1$   \\ [0.5ex]
\hline
\end{tabular}}
\end{center}
\caption{The first row shows the control parameters of the three kinds of waves considered here: the Froude number, $Fr$, and the Weber number, $We$, describe the competition between the shear in the air and the relevant restoring force; $\cmd$ and $\kcd$ are a dimensionless minimum phase speed and a dimensionless capillary wavenumber, respectively. The second row gives the corresponding dimensionless dispersion relations. The third row gives the small parameter, $m\ll1$, defining the strong wind limit for each, and the last row gives the exponents $q$ characterizing the associated asymptotic states in the case of the exponential wind profile. }
\label{tablewave}
\end{table}
The wind profile has velocity scale $V$, and length scale $L$, giving the dimensionless variables
\refstepcounter{equation}
$$
  \z= \frac{z}{L},\quad\kd = kL, \quad\U= \frac{U}{V}, \quad\text{and}\quad \C = \frac{c_0}{V}.
  \eqno{(\theequation{\mathit{a},\mathit{b},\mathit{c},\mathit{d}})}
$$
We consider two standard wind profiles, shown in Figure \ref{schema}. For the exponential profile, $V=U_{\infty}$ and $L=d$, and for the logarithmic profile, $V=u_{\star}$ and $L=z_0$, where all symbols are defined in the legend of Figure \ref{schema}. Their dimensionless forms are
\refstepcounter{equation}
$$
\U(\z)=1-e^{-\z}\qquad\text{and}\qquad \U(\z)= \ln(1+\z)/\kappa,
  \eqno{(\theequation{\mathit{a},\mathit{b}})}
$$
respectively. We stress that typical values of $z_0$ are of the order of millimetres \citep{wu75} while the wavelengths of capillary--gravity waves range from millimetres to hundreds of metres. Thus, most wind waves are long in the sense that their wavelengths are much greater than the characteristic length scale of the wind profile, and $\kd= kz_0$ is a natural small parameter.

For the three dispersion relations given in Table \ref{tablewave}, we solve the following boundary-value problem:
\begin{widetext}
\refstepcounter{equation}
$$
\chi''(\z) -\bigg[\kd^2 + \frac{ \U''(\z)}{\U(\z)-\C(\kd)}\bigg]\chi(\z)=0,\qquad \chi(0) = 1, \qquad\chi'(\z) +\kd\ \chi(\z) \underset{\z\to+\infty}{\longrightarrow} 0 \label{solved}
  \eqno{(\theequation{\mathit{a},\mathit{b},\mathit{c}})},
$$
\end{widetext}
where $\chi\equiv \hat{\psi}_0/\hat{\psi}_0(0)$ is the leading-order normalized streamfunction amplitude. We emphasize that physically relevant wind profiles satisfy
\be
\lim\limits_{\z\to+\infty}\frac{ \U''(\z)}{\U(\z)-\C(\kd)} = 0, 
\label{eq:just}
\ee
which justifies the far field condition (\ref{solved}\textit{c}).
In practice, the coefficients defined in equation (\ref{phillips}), which are more physical than the $\alpha$ and $\beta$ introduced by \citet{miles57}, are calculated as follows:
\refstepcounter{equation}
$$
\mu =\epsilon \ \bigg(\frac{\U'}{\kd\C} +\frac{ \Re\{\chi'\}}{\kd}\bigg) \bigg|_{0^+}\quad\text{and}\quad  \gamma = \frac{\epsilon}{\kd}\  \Im\big\{\chi'(0^+)\big\}.\label{coeffs}
  \eqno{(\theequation{\mathit{a},\mathit{b}})}
$$
Note that $\alpha=\epsilon \mu/\C^2$ and $\beta=\epsilon \gamma/\C^2$. The Miles formula states that \citep{janssenbook} 
\be
\gamma = -\epsilon\ \frac{\pi}{\kd}\ \frac{\Uppc}{\Upc}\ |\crit|^2, \label{miles}
\ee
where the subscript `c' denotes evaluation at the critical level $\zcd = \zcd (\kd)$, defined by 
\be
U(\zcd)=\C.
\ee
The expression (\ref{miles}) originates from the global property of the solution of the boundary-value problem (\ref{solved}), 
\be
\Im\big\{\chi'(0^+)\big\} =-\pi\ \frac{\Uppc}{\Upc}\ |\crit|^2,\label{property}
\ee
which we use to assess the accuracy of our numerical solutions. 

We evaluate the accuracy of the asymptotic methods developed here using the asymptotic suction boundary layer profile, $\U(\z) =1-e^{-\z}$, for which an exact solution of the Rayleigh equation exists \citep{young-wolfe}. However, for comparison with experimental data we shall use the more common mean turbulent boundary layer profile, $\U(\z) =\ln(1+\z)/\kappa$. 

\section{Long wave asymptotics} \label{longwaves}

Long waves are characterized by $\kd\ll1$. \textcolor{black}{The following analysis is valid for the three functions $\C(\kd)$ given in Table \ref{tablewave}. For capillary--gravity waves, the wavelength is of the order of the capillary wavelength, $\lc\equiv 2\pi/ \kc$, so $\lc$ must be large compared with $L$, namely
\be
\kcd\ll1. \label{longcapgrav}
\ee 
\subsection{General procedure}}
Setting $\kd=0$ in equation (\ref{solved}\textit{a}), we find two linearly independent solutions \citep{drazin-reid}, 
\refstepcounter{equation}
$$
\chi_1(\z) \equiv \U(\z)-\C \quad\text{and}\quad \chi_2(\z) \equiv \chi_1(\z) \int^\z \frac{d\tilde{z}}{\chi_1(\tilde{z})^2}\ . 
  \eqno{(\theequation{\mathit{a},\mathit{b}})} 
$$
We call the outer solution the linear combination of $\chi_1$ and $\chi_2$, namely
\be
\out(\z)\equiv E\ \chi_1(\z) + F\ \chi_2(\z),\quad\text{with}\quad E,F\in \mathbb{C}.
\ee
\textcolor{black}{We consider wind profiles such that $\U'>0$, $\U''<0$ and $\U'''>0$, so that there exists a unique position} $\zs$ between the critical level, $\zcd$, and infinity at which
\be
Q(\kd,\zs) = 0,\quad\text{where}\quad Q(\kd,\z)\equiv\kd^2+\frac{\U''(\z)}{\U(\z)-\C(\kd)}\ . \label{condi}
\ee
\textcolor{black}{Then the outer solution holds for $\z\ll\zs$. Eq. \eqref{eq:just} together with the far field condition (\ref{solved}\textit{c}) imply that $\chi(\z)\sim \solinf(\z)$ for $\z\gg\zs$, where
\be
\solinf(\z)\simeq G\ e^{-\kd\z},\qquad\text{with}\qquad G \in\mathbb{C}. \label{farfield}
\ee
We stress that $\zs=\zs(\kd)$. For standard wind profiles, we show in Appendix \ref{Appinflex} that 
\be
\lim\limits_{\kd\to0} \zs =+\infty,\qquad\text{but}\qquad \lim\limits_{\kd\to0} \kd\zs =0. 
\ee
In order to match the outer solution and the far field solution within an intermediate layer centred at $\z=\zs$, we introduce the rescaled variable $\xi\equiv \kd \z$. Then, we determine the constants $E$, $F$ and $G$ using the matching condition 
\be
\lim\limits_{\z\to+\infty} \out(\z) = \lim\limits_{\xi\to0} \solinf(\xi). \label{match}
\ee}
Clearly, the asymptotic behaviour of $\out$ depends on the choice of $\U=\U(\z)$, whereas 
\be
\solinf(\xi)\sim G(1- \xi),\qquad \xi\to0.
\ee
Hence, there are profiles, such as the logarithmic profile, for which matching is not possible. However, we note that the solution of the Rayleigh equation has an inflexion point at $\z=\zs$, and thus its behaviour is linear within the intermediate layer. Therefore, we anticipate that patching, rather than rigorous matching, of $\out$ and $\solinf$ at $\z=\zs$ will  still give reasonable results.  \textcolor{black}{In order for rigorous matching of solutions in all cases, a more detailed treatment around the point $\z=\zs$ is necessary, but we provide numerical evidence that the present approach faithfully reproduces the behaviour in this region.}

\textcolor{black}{In practice, we work with a transformed variable $\Z = \Z(\z,\zcd)$, such that the function $Q$ introduced in equation (\ref{condi}) becomes independent of $\C(\kd)$. Using this transformation, the domain $[0,+\infty[$ becomes $[\Z_{\rm{inf}},+\infty[$, where $\Z_{\rm{inf}}= \Z_{\rm{inf}}(\zcd)$ depends on the wind profile and can be negative. In all cases considered here, we check that $\Z_{\rm{inf}} \le 1$, even in the limit $\kd\to 0$. 
\subsection{Matching for the exponential profile: $\U(\z) =1-e^{-\z}$}
For this profile, we use the variables
\refstepcounter{equation}
$$
\Z \equiv \z-\zcd  \qquad\text{and}\qquad \X(\Z)\equiv\chi(\z), 
  \eqno{(\theequation{\mathit{a},\mathit{b}})}
$$ 
in terms of which the boundary-value problem (\ref{solved}) becomes
\be
\X''(\Z)- \Bigg[\kd^2+ \frac{1} {  1-e^{\Z}   }  \Bigg]  \ \X(\Z)=0,
\ee
\refstepcounter{equation}
$$
\X(-\zcd)=1,\qquad\X'(\Z) +\kd\ \X(\Z) \underset{\Z\to+\infty}{\longrightarrow} 0.  \label{BCexp}
  \eqno{(\theequation{\mathit{a},\mathit{b}})}
$$ 
Here, the outer solution is 
\be
\X_{\rm{out}}(\Z) =E (1-e^{-\Z}) + F \big(1-e^{-\Z} \big) \bigg\{\frac{1}{1-e^\Z}+\Log\big(e^\Z-1\big)\bigg\},
\ee
where $\Log$ denotes a continuation of the natural logarithm to the negative real numbers
\be
\Log(\z-\zcd)\equiv \ln|\z-\zcd|- i\qquad \text{for }\z<\zcd. 
\label{Log}
\ee
The choice of the branch cut, which is just above the negative real axis as $\Upc>0$, follows from \citet{linbook}. The matching condition (\ref{match}) gives $E=G$ and $F=-\kd\ G$, with the remaining constant, $G$, being determined by the boundary condition (\ref{BCexp}\textit{a}).} 

\textcolor{black}{We construct a uniformly valid composite solution using the Van Dyke additive rule \citep[see e.g.,][]{B-O} as}
\begin{widetext}
\begin{align}
\Xu(\Z)&= \Xout(\Z)+\Xinf(\Z)-(E+F\ \Z)\nn \\
&= G(\kd,\pp)\ \bigg\{ 1- e^{-\Z} - \kd \bigg[  \frac{1-e^{-\Z}}{1-e^{\Z}} + \big(1-e^{-\Z}\big)\Log\big( e^{\Z} -1\big) \bigg] + e^{-\kd\Z} - (1-\kd\Z) \bigg\}, \label{unifexp}
\end{align}
\end{widetext}
where
\be
G(\kd,\pp) = \frac{1-\pp}{1-\kd \pp + \kd\ln(\pp) + i\kd\pi}\ , \quad\text{and}\quad \pp\equiv\C^{-1}. \label{Gexp}
\ee
Note that $\pp = \pp(\kd)$ because of the dispersion relation, $\C = \C(\kd)$. In Appendix \ref{Appexp}, we retrieve the expression for $\crit$ when $\zcd \ll1$ that Miles derived in an Appendix to \citet{morland-saffman}. 
\subsection{Patching for the logarithmic profile: $\U(\z) =\ln(1+\z)/\kappa$}
We perform the coordinate transformation
\refstepcounter{equation}
$$
\Z \equiv \frac{1+\z}{1+\zcd}  \qquad\text{and}\qquad \X(\Z)\equiv\chi(\z), \label{newvar'}
  \eqno{(\theequation{\mathit{a},\mathit{b}})}
$$ 
and hence the boundary-value problem (\ref{solved}) becomes
\begin{widetext}
\refstepcounter{equation}
$$
\X''(\Z)- \Bigg[\Long^2- \frac{1} {  \Z^2 \ln(\Z) }  \Bigg]   \ \X(\Z)=0,\qquad \X\bigg(\frac{1}{1+\zcd}\bigg)=1,\qquad\X'(\Z) +\Long\ \X(\Z) \underset{\Z\to+\infty}{\longrightarrow} 0,
  \eqno{(\theequation{\mathit{a},\mathit{b},\mathit{c}})}
$$ 
where we have introduced $\Long\equiv \kd(1+\zcd)$. We use $\Long$ instead of $\kd$ as a small parameter. After a patching at $\Zs \equiv (1+\zs)/(1+\zcd)$, we find the following outer solution:
\be
\Xout(\Z) = 
\bc
\Big( J(\Zs) \ln(\Z)-H(\Zs)\big[\ln(\Z)\li(\Z)-\Z\big]\Big)  \frac{G(\zcd,\Zs)}{\Zs }\ e^{-\Long \Zs} \qquad\text{if }\Z>1,\\
\\
\Big( J(\Zs) \ln(\Z)-H(\Zs)\big[\ln(\Z)\big[\li(\Z) -i \pi\big]-\Z\big]\Big)  \frac{G(\zcd,\Zs)}{\Zs }\ e^{-\Long \Zs} \qquad\text{if }\Z<1,
\ec \label{outerlog}
\ee
\end{widetext}
where
\be
\li(\Z)\equiv \cauchy\int_0^\Z \frac{d\tilde{z}}{\ln(\tilde{z})}
\ee
is the logarithmic integral function, in which $\cauchy$ denotes the Cauchy principal value. The amplitude of the far field solution is
\be
G(\zcd,\Zs) = \frac{\Zs(1+\zcd)e^{\Long \Zs}}{H(\Zs) g(\zcd)-J(\Zs) f(\zcd)-i\pi H(\Zs) f(\zcd)}\ , \label{Glog}
\ee
where
\begin{align}
H(\Zs) &\equiv \frac{1}{\Long \Zs}+1, \\
 J(\Zs)&\equiv H(\Zs)\li(\Zs)-\Long \Zs^2\ ,\\
f(\zcd)&\equiv(1+\zcd)\ln(1+\zcd),\qquad\text{and} \\
g(\zcd)&\equiv1+f(\zcd)\li\bigg(\frac{1}{1+\zcd}\bigg).
\end{align}
Clearly, for a given dispersion relation, the above parameters are all functions of $\kd$. \textcolor{black}{Matching is not possible here because the behaviour of $\Xout(\Z)$ at large $\Z$ is not linear.
\subsection{Discussion} \label{discussion}
\begin{figure*}[htbp!]
        (a)\includegraphics[trim = 0 0 0 0, clip, width = 0.47\textwidth]{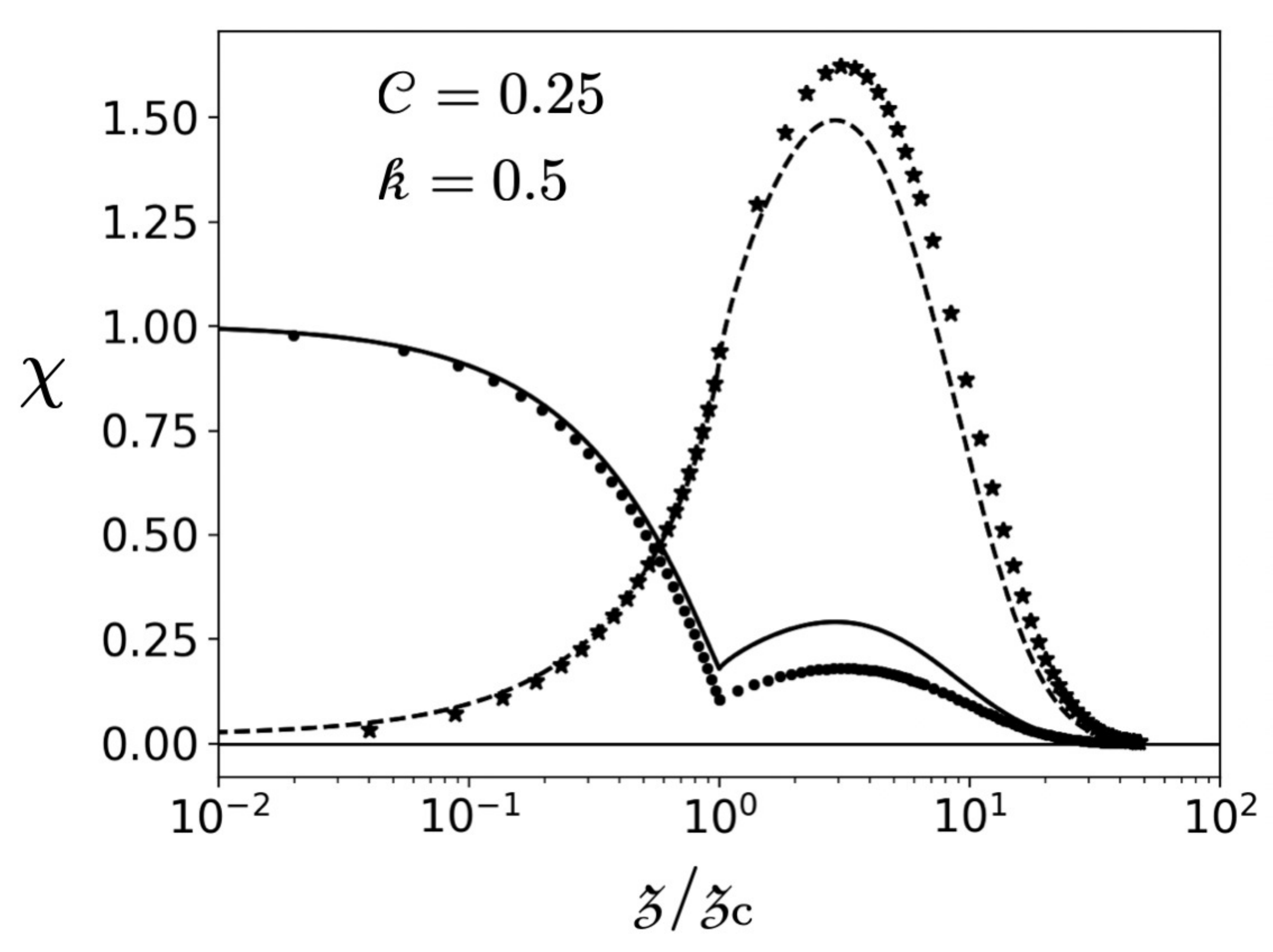}	  	
        (b)\includegraphics[trim = 0 0 0 0, clip, width = 0.45\textwidth]{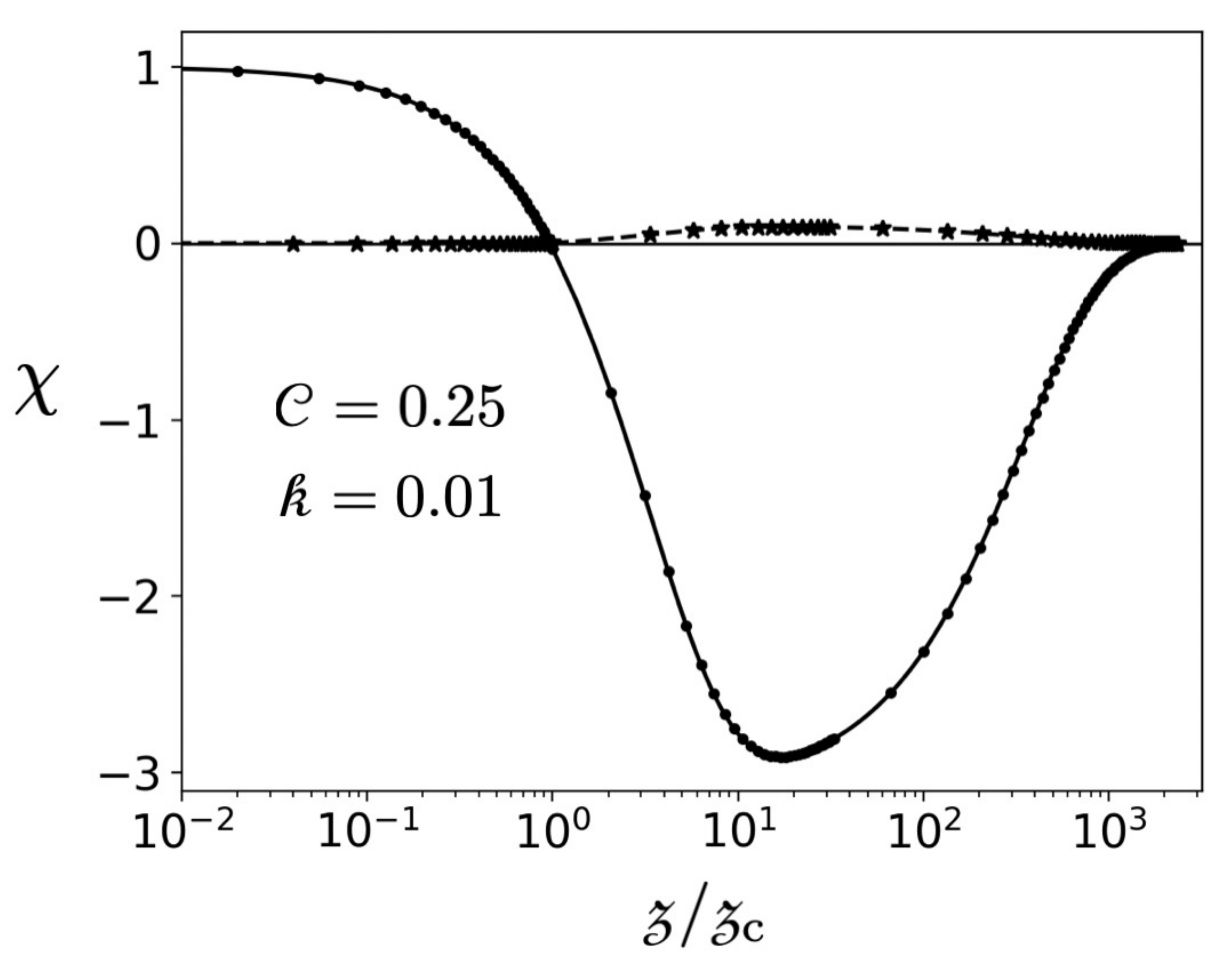}
    \caption{Comparison of the uniformly valid composite solution (\ref{unifexp}) with the numerical solution of the Rayleigh equation for the exponential wind profile, for two values of $\kd$ and $\C$=0.25. The dots and the stars depict the real and imaginary parts of the numerical solution, respectively. The continuous line shows the real part of (\ref{unifexp}) and the dashed line the imaginary part.}
\label{longexp}
\end{figure*}
\begin{figure*}[htbp!]
        (a)\includegraphics[trim = 0 0 0 0, clip, width = 0.46\textwidth]{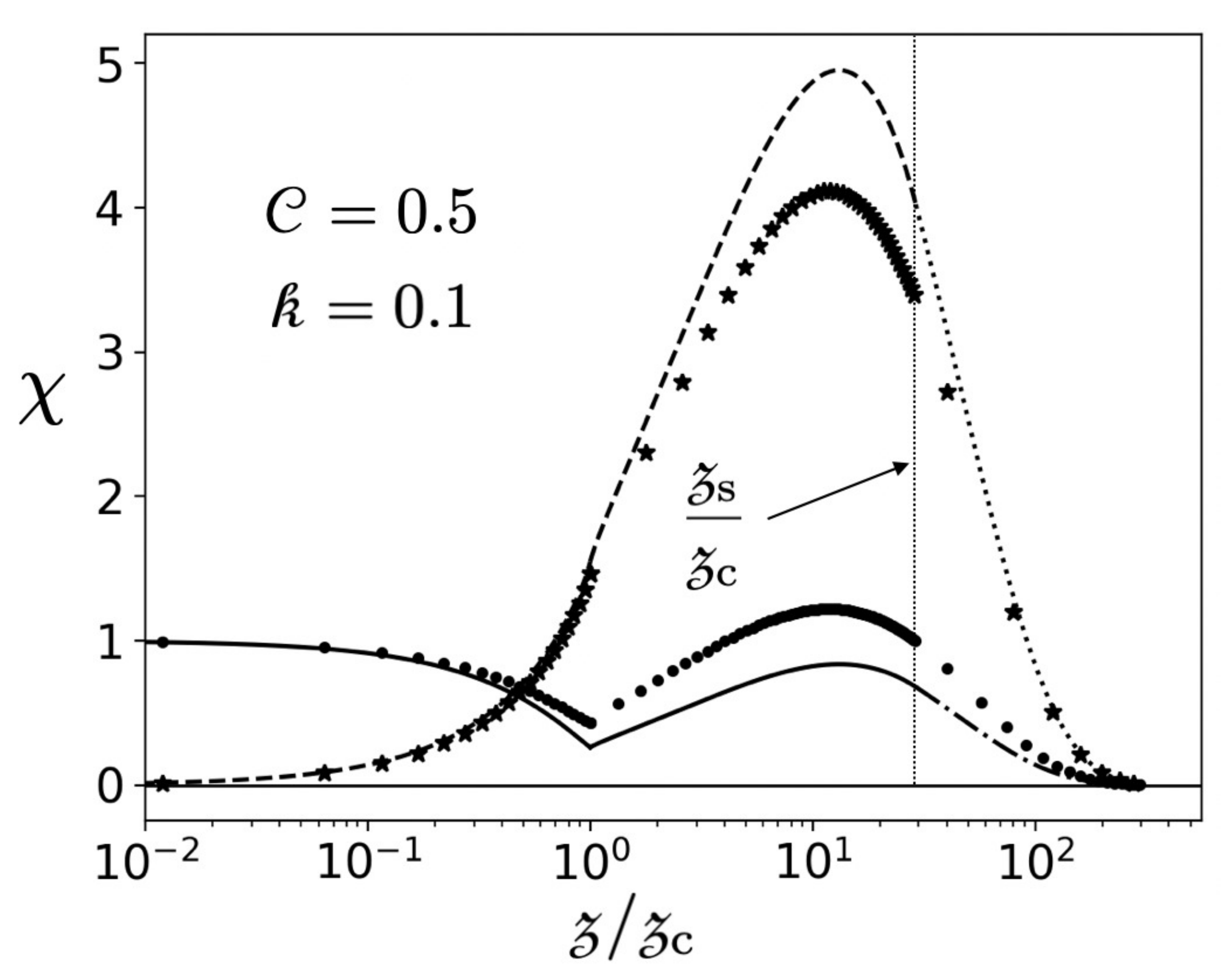}	  	
        (b)\includegraphics[trim = 0 0 0 0, clip, width = 0.47\textwidth]{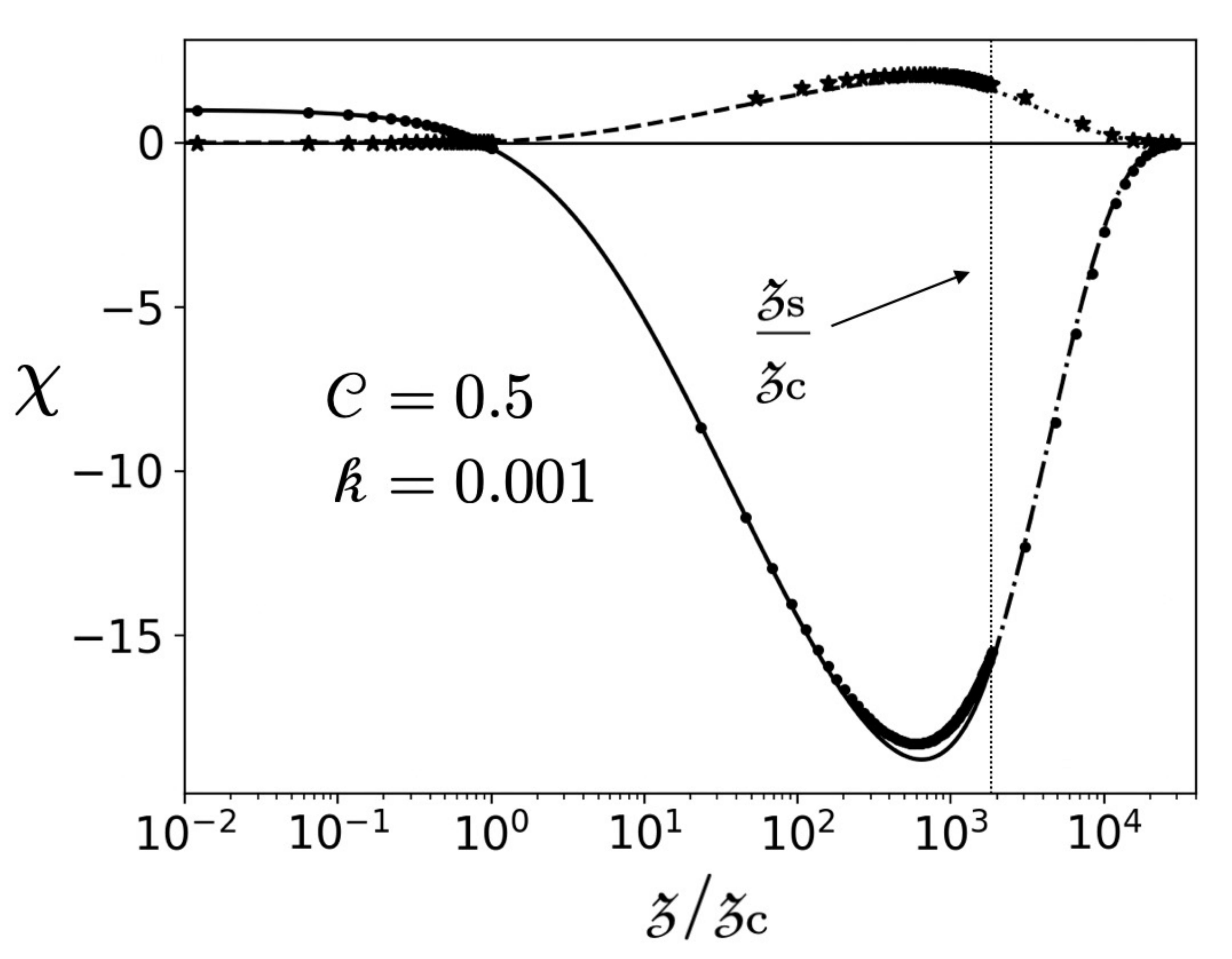}
    \caption{Comparison of the outer and far field solutions patched at the inflexion point, $\z=\zs$, with the numerical solution of the Rayleigh equation for the logarithmic wind profile, for two values of $\kd$ and $\C$=0.5. The dots and the stars depict the real and imaginary parts of the numerical solution, respectively. The continuous line shows the real part of the outer solution (\ref{outerlog}) and the dashed line the imaginary part. The dash-dotted and dotted lines represent the real and imaginary parts of the far field solution (\ref{farfield}), respectively.}
\label{longlog}
\end{figure*}
In Figures \ref{longexp} and \ref{longlog}, we} compare our uniformly valid composite solution for the exponential profile, and our patched solution for the logarithmic profile with the numerical solutions. \textcolor{black}{Both the matching and the patching give excellent results for sufficiently small values of $\kd$, the magnitude of which depends on the wind profile.  We note that these are plots for fixed values of $\kd$ and $\C$ and that any of three dispersion relations can be retrieved with a proper choice of the control parameters.} Moreover, we assess our approach by checking that $\Xu(\Z)$ and $\Xout(\Z)$ satisfy the global property (\ref{property}). Above the critical level, the phase of the solution of the Rayleigh equation is constant, equal to the phase of $G$, showing that long waves interact with the wind between the mean water level and the critical level. 

For the two standard wind profiles considered here, both the real and the imaginary part of the solution of (\ref{solved}) have an extremum at $\z=\zex$, between the critical level, $\zcd$, and the inflexion point, $\zs$. In Appendix \ref{Appfixedpoint}, we show the extremum is always a maximum for the imaginary part, whereas for the real part it is a minimum when $\kd<\kd_{\star}$ but a maximum when $\kd>\kd_{\star}$; where $\kd_{\star}$ is the wavenumber of the fastest growing wave. We also show that the air flow above wind waves has two elliptic points at the level $\z=\zex$ in the domain $kx\in[0,2\pi[$. These elliptic points can be seen in Figure 13a of \citet{young-wolfe}, obtained from the hypergeometric solution of the Rayleigh equation in the case of the exponential profile, and in Figure 1(e) of \citet{hristov-nature}, obtained from the numerical solution for the logarithmic profile.
\section{Application to the Miles instability}\label{appli}
\subsection{Normalized energy growth rate and deviation from the sheltering hypothesis} \label{appli1}
\begin{figure*}[htbp!]
        (a)\includegraphics[trim = 0 0 0 0, clip, width = 0.49\textwidth]{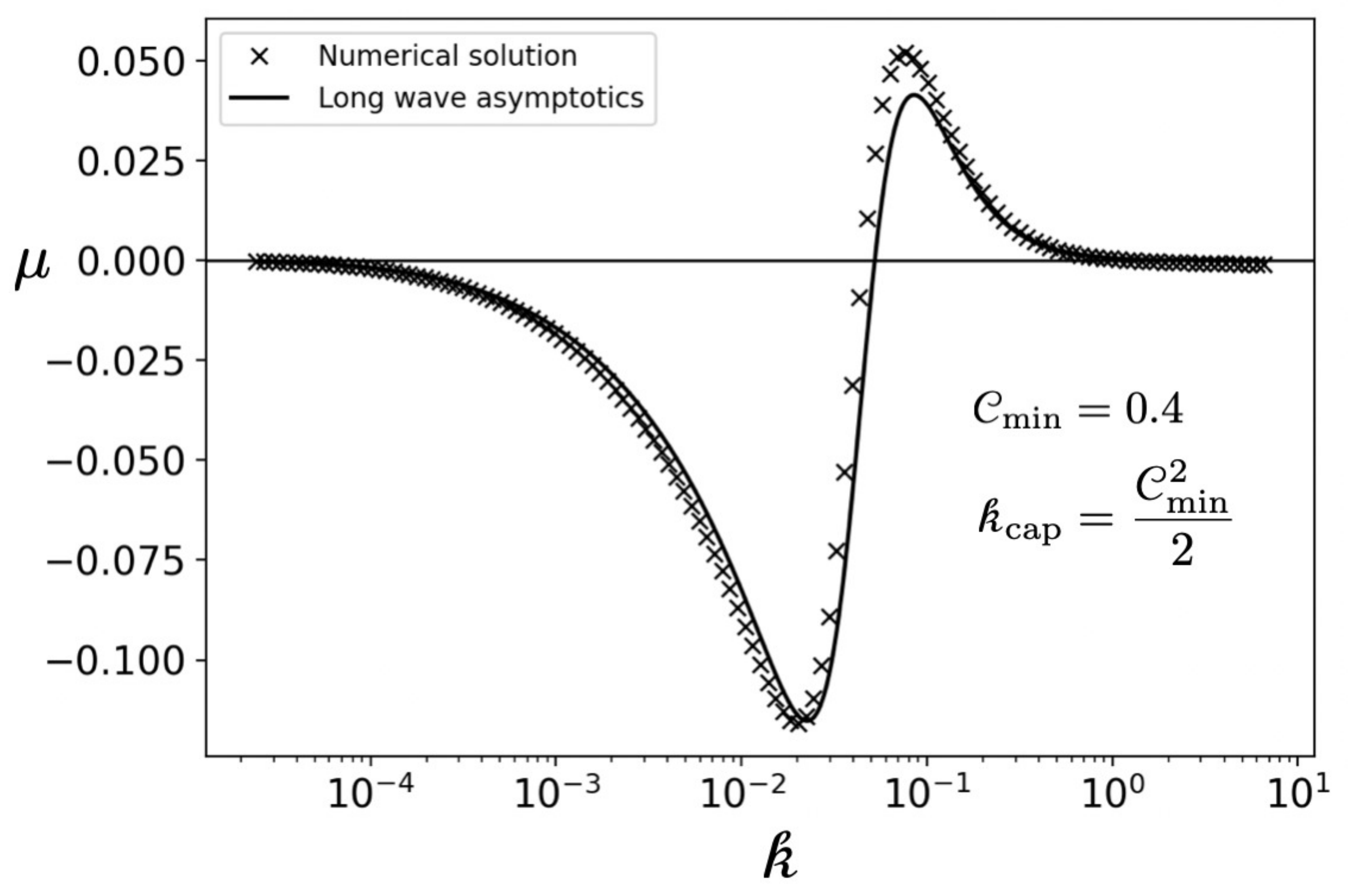}	  	
        (b)\includegraphics[trim = 0 0 0 0, clip, width = 0.45\textwidth]{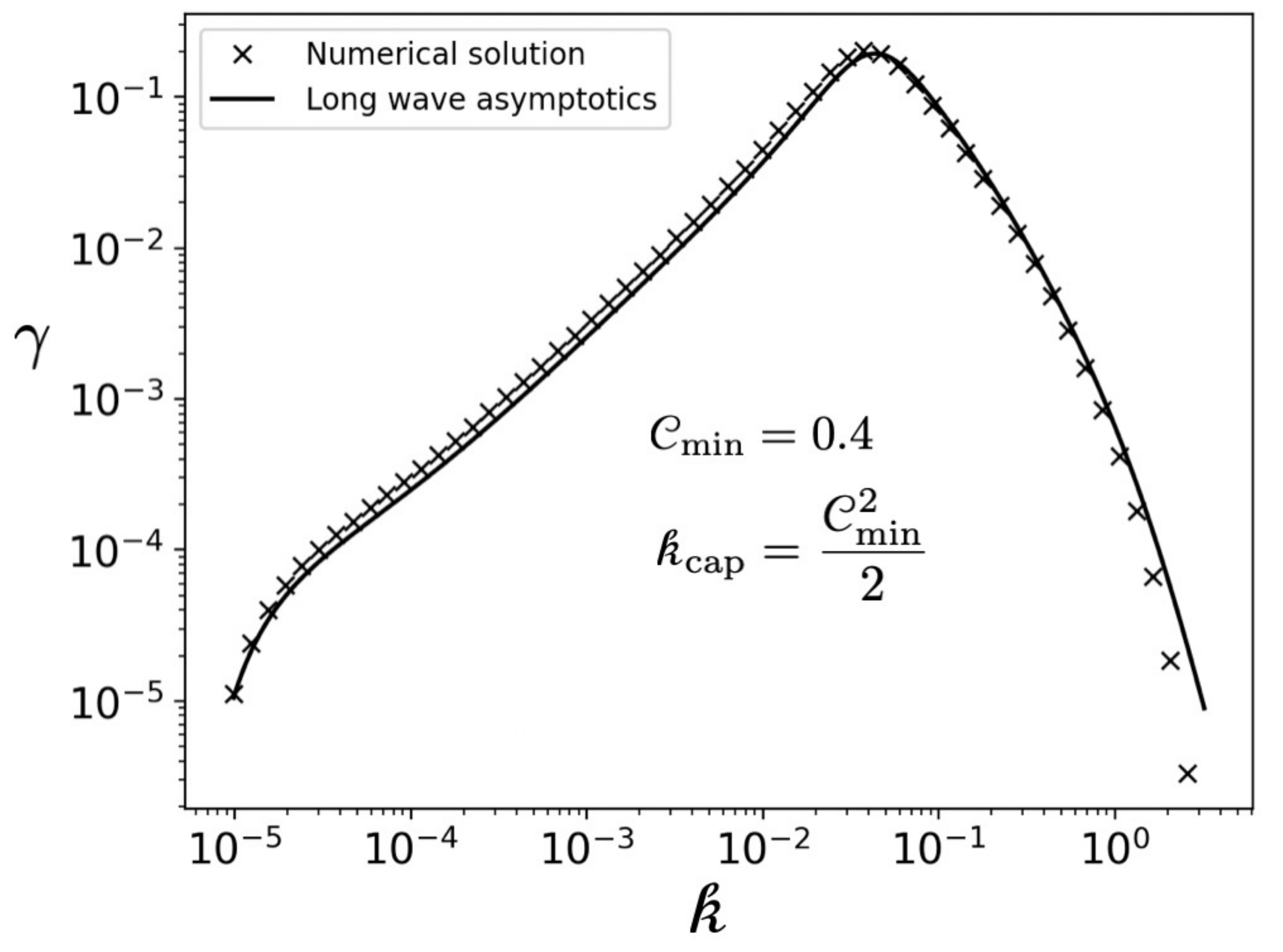}
        (c)\includegraphics[trim = 0 0 0 0, clip, width = 0.75\textwidth]{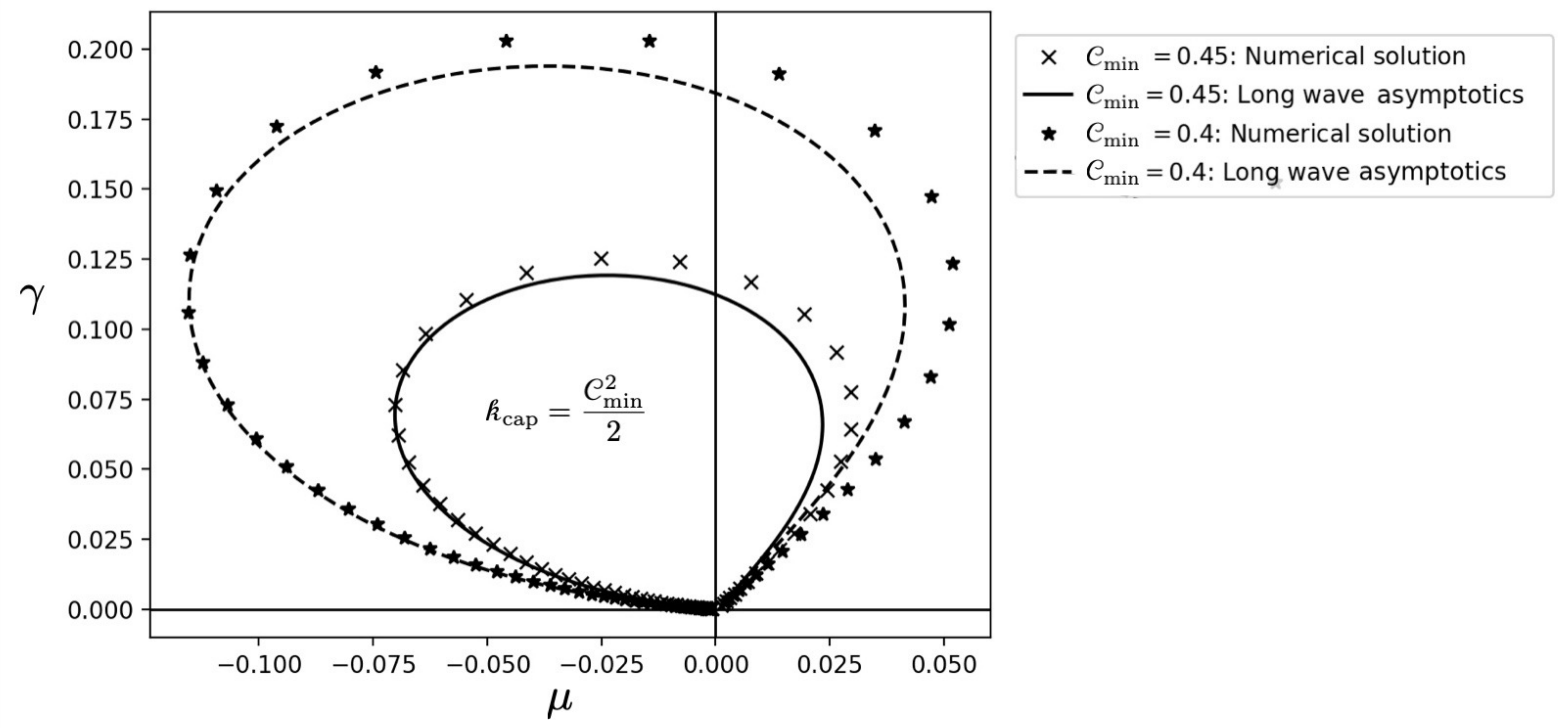}
 \caption{Long wave asymptotic results for capillary--gravity waves and the logarithmic profile. (a) Twice the wind-dependent relative change of phase speed, $\mu$. (b) The normalized energy growth rate, $\gamma$, as a function of the dimensionless wavenumber, $\kd = kL$, where $L$ is the length scale associated with the wind profile. (c) Plot of $\gamma$ versus $\mu$ for two values of $\cmd.$ }
\label{logmugamma}
\end{figure*}
We calculate the coefficients $\mu$ and $\gamma$ for long waves using the expressions (\ref{coeffs}\textit{a,b}). In the case of the exponential profile, we find
\begin{align}
\mu^{\rm{exp}}_{\rm{long}} (\kd)&= -\epsilon\ \frac{(\pp-1)^2[1-\kd \pp + \kd\ln(\pp)]}{[1-\kd \pp + \kd\ln(\pp)]^2 + [\kd \pi]^2 }\qquad  \text{and}\nn \\
\gamma^{\rm{exp}}_{\rm{long}}(\kd) &= \frac{\epsilon \ \pi\ \kd (\pp-1)^2 } {[1-\kd \pp + \kd\ln(\pp)]^2 + [\kd \pi]^2 }\ , \label{gammaexp}
\end{align}
and in the case of the logarithmic profile, we find
\begin{figure}
\centering
\includegraphics[ width=\columnwidth]{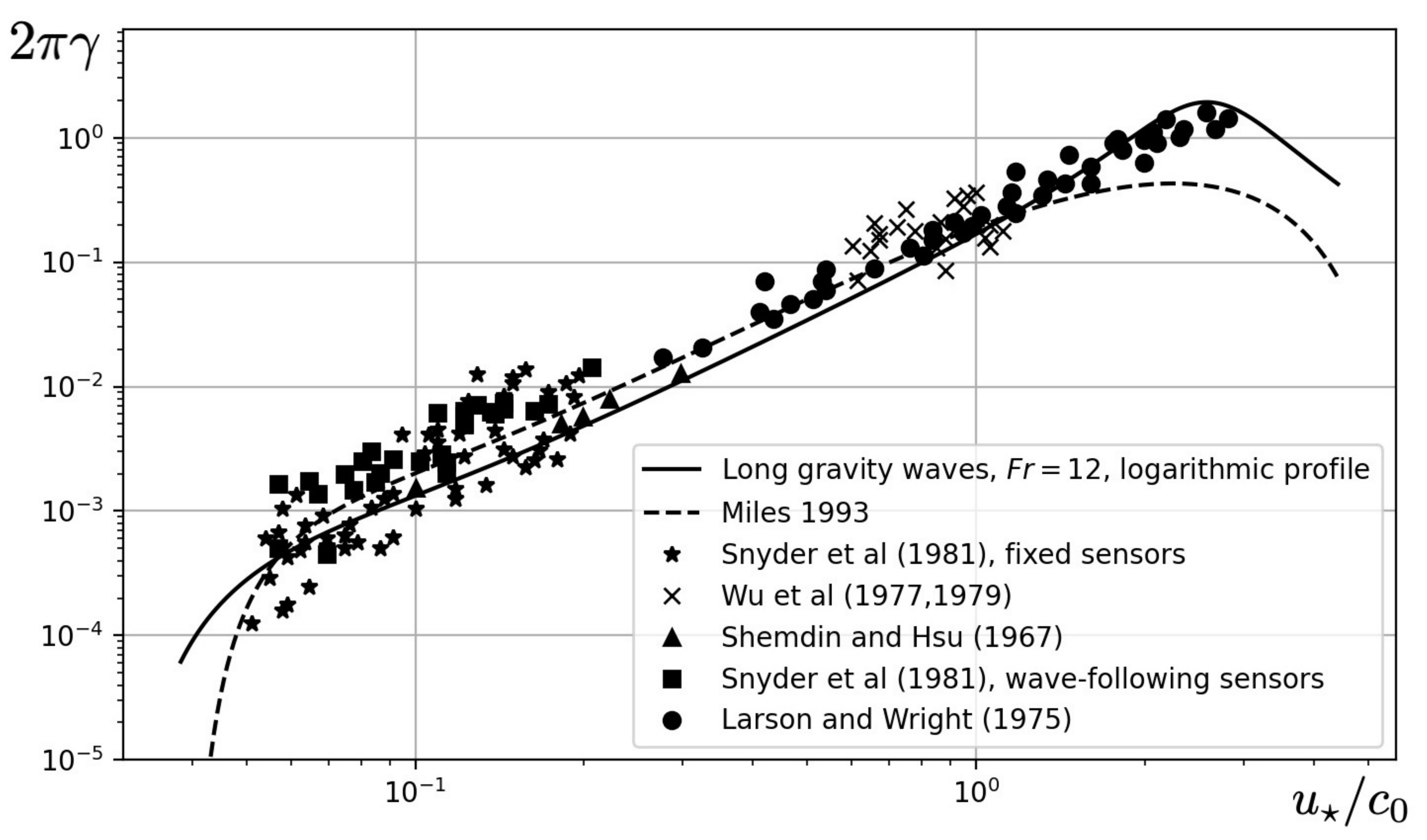}
  \caption{Comparison of the normalized energy growth rate (multiplied by $2\pi$) calculated using the long wavelength asymptotics for the logarithmic profile and gravity waves characterized by a Froude number $Fr=12$, with the experimental data compiled by \citet{plant82}. The dashed line shows the results of \citet{miles93} for the same Froude number.  }
\label{plantmiles}
\end{figure}
\begin{widetext}
\begin{align}
\mu^{\rm{log}}_{\rm{long}}(\kd) &= \frac{ \epsilon H(\Zs)}{\kd \ln(1+\zcd)}\ \frac{ H(\Zs)g(\zcd) -J(\Zs)f(\zcd)}{\big[ H(\Zs)g(\zcd) -J(\Zs)f(\zcd)\big]^2 +\big[\pi H(\Zs) f(\zcd) \big]^2}\ \qquad \text{and}\nn \\
\gamma^{\rm{log}}_{\rm{long}}(\kd) &= \frac{\epsilon}{\kd}\ \frac{\pi (1+\zcd) H^2(\Zs)}{\big[ H(\Zs)g(\zcd) -J(\Zs)f(\zcd)\big]^2 +\big[\pi H(\Zs) f(\zcd) \big]^2}\ . \label{gammalog}
\end{align}
\end{widetext}
The dependence upon the physical parameters, $Fr$, $We$, $\cmd$ and/or $\kcd$, is contained in the inverse phase speed, $\pp$, for the exponential profile, or the critical level, $\zc$, and the transformed inflexion point, $\Zs$, for the logarithmic profile. 

For capillary--gravity waves and the logarithmic profile  we compare the numerical evaluation of $\mu$ and $\gamma$ with our asymptotic expressions in Figure \ref{logmugamma}, and note that the plots for the exponential profile are very similar. In anticipation of the strong wind limit (see \S \ref{logSW}), we have chosen the control parameters, $\kcd$ and $\cmd$, such that the fastest growing waves are driven by both gravity and surface tension. We maintain the scaling of the wavenumber with the length scale of the wind profile, $L$, although we note that we could also use the capillary wavelength. For both profiles, the asymptotics show very good agreement with the numerics, even when $\kd =O(1)$. The normalized growth rate, $\gamma$, has a maximum at $\kd =\kmax$ in the long wave regime. The deviation from the Jeffreys sheltering hypothesis, as captured by $\mu$ (cf. \ref{phillips}), is equal to zero for a wavenumber close to $\kd =\kmax$, indicating that the fastest growing wave is such that the aerodynamic pressure is almost in phase with the wave slope. Thus, we demonstrate the validity of Jeffreys' intuition of wind-wave growth and show that the assumption of \citet{conte-miles} and \citet{milespart2} that $\alpha<0$ was erroneous.

\subsection{Classical case: logarithmic profile and $\sigma=0$}

\citet{plant82} collected experimental data for the normalized energy growth rate (multiplied by $2\pi$). In Figure \ref{plantmiles}, we compare his results with the long wave asymptotics for the logarithmic profile and gravity waves characterized by a Froude number $Fr=12$\textcolor{black}{; the range of $\kd$ used here is $[10^{-5}, 0.135]$. }Our analysis provides a good fit of the entire range of data, contrary to that of \citet{miles93}. Nonetheless, the measurements were made in different conditions and the data analysed using different dispersion relations; for instance, \citet{larson-wright} considered capillary--gravity waves. Therefore, it would be more appropriate to consider a range of Froude numbers, or more generally a range of $\cmd$ and $\kcd$, the control parameters for capillary--gravity waves. In addition to the challenging aspects of making these measurements, this may explain the significant scatter of the data.

\subsection{Interpretive framework of the Miles mechanism}

\begin{figure}
  \centerline{\includegraphics[width=1.1\columnwidth]{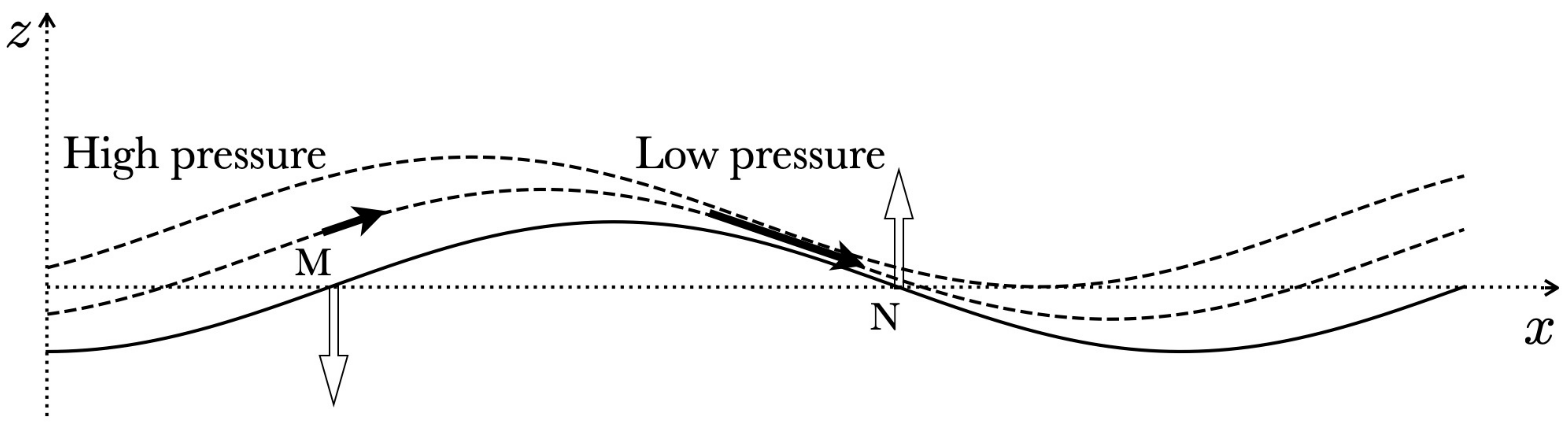}}
  \caption{Streamlines of the air flow (dashed) over water waves, modified from \citet{stewart74}. The solid line depicts the water surface, defined by the displacement $\eta(x,t) = a\cos(kx-\omega t)$. The wave slope is $\del_x\eta(x,t) =- ka\sin(kx-\omega t)$, and the vertical speed of the points on that curve is $\del_t\eta(x,t) = \omega a\sin(kx-\omega t)$. Point $M$ has a phase equal to $-\pi/2$ (positive slope), and thus moves downward. Point $N$ has a phase equal to $\pi/2$ (negative slope), and thus moves upward. The pressure asymmetry, caused by the phase shift of the streamlines, enhances the motion of $M$ and $N$. Thick black arrows represent the velocity field of the air flow perturbation, $\bb{u}=u\ \bb{\hat{x}}+ w\ \bb{\hat{z}}$. We observe that $\overline{uw}<0$, where the overbar denotes the average over a wavelength. }
\label{mechanism}
\end{figure}

\citet{jeffreys25} proposed that wind waves grow because of a pressure asymmetry due to flow separation: the air flowing over a wave separates on the downwind side and reattaches on the upwind side of the next crest. \citet{banner-melville} argue that there is no air flow separation unless the waves break.  In \ref{appli1} we showed that the condition for optimal growth is equivalent to 
Jeffreys' idea of the aerodynamic pressure being in phase with the wave slope.   We develop this finding in section \ref{strongwind} with the aid of the strong wind limit. Jeffreys' idea can be understood as follows. In the absence of wind, the aerodynamic pressure is in a phase opposite to that of the surface displacement -- high pressure at the wave troughs, low pressure at the wave crests -- and the streamlines of the air flow adjust to the water surface. For growth to happen, a phase shift of the streamlines is required \citep{lighthill62,stewart74}, as shown in Figure \ref{mechanism}. The optimal phase shift can even be intuited by observing that a node on the windward side (point $M$) is moving down while a node on the leeward side (point $N$) is moving up. Therefore, if the pressure is maximal windward and minimal leeward, the motion of the nodes will be enhanced, and hence the optimal phase shift is equal to $\pi/2$. 

Mathematically, a non-zero phase shift can only arise from a complex leading-order streamfunction amplitude;  $\hat{\psi_0}(z)\in \mathbb{C}$. We recall that $\hat{\psi_0}$ is the first term in an expansion in powers of the air/water density ratio $\epsilon$, regarded as a coupling constant. Therefore, $\hat{\psi_0}$ determines the neutral perturbation induced by the water waves on the air flow. Because the Rayleigh equation (\ref{L_0}) has real coefficients, only a singularity can lead to a complex solution and hence to a critical layer.

In \S \ref{model}, we introduced the wave-induced Reynolds stress, $\tau=-\dair\ \overline{uw}$, where $u$ and $w$ are the velocity components of the air flow perturbation and the overbar denotes the average over a wavelength. It is evident from Figure \ref{mechanism} that a phase shift of the streamlines implies that $\tau >0$. This view of the growth in terms of a positive wave-induced Reynolds stress was first pointed out by \citet{stewart74} and formalized by \citet{janssenbook}, and we made the connection with the energy growth rate in the end of \S \ref{perturbationtheory}. The leading-order amplitude of $\tau(z,t)$ is shown in Eq. \eqref{tau0} and clearly displays the necessity of a complex streamfunction. (Note that the quantity on the right hand side of  Eq. (\ref{tau0}) is piecewise constant.) As a consequence of the global property (\ref{property}), $\hat{\tau}_0$ maintains the same positive value from the water surface up to the critical level, $z=z_c$, and then jumps to zero as it must vanish in the far field. 

Since the original work of \citet{miles57} our understanding of the basic question of how wind waves grow has advanced in three key aspects:
(i) waves grow due to the work of a positive wave-induced Reynolds stress; (ii) waves grow due to the phase shift of the streamlines of the air flow; and (iii) waves grow due to an asymmetric pressure distribution. Our contribution here is to prove that for optimal growth the phase shift must be equal to $\pi/2$, as intuited by Jeffreys. Furthermore, we made clear that all these contributions have the same mathematical basis -- a complex streamfunction -- and the same physical foundation -- a critical layer. 
\section{Strong wind limit} \label{strongwind}
Following \citet{young-wolfe}, we introduce a parameter $m$, which is a measure of the strength of the wind. As seen in Table \ref{tablewave}, $m$ depends on the restoring force. In the strong wind limit, defined by $m\ll1$, $\kmax$ tends to the point at which $\mu_{\rm{long}}$ vanishes, which shows that the Jeffreys sheltering hypothesis is in fact the condition for optimal growth of wind waves. A derivation of the results stated below is given in Appendix \ref{Appmax}.
\textcolor{black}{\subsection{Exponential wind profile} \label{SWexp}
For $\U(\z)=1-e^{-\z}$, when $m\ll1$ }the normalized energy growth rate becomes a Lorentzian function
\be
\frac{\gamma^{\rm{exp}}_{\rm{long, SW}}(\kd)}{\gm(q)} =   \frac{\big[\Delta(q)\big]^2 }{\big[\kd-\kd_{\star}\big]^2+ \big[\Delta(q)\big]^2}\ , \label{gammaSW}
\ee
where `SW' denotes `strong wind', and
\refstepcounter{equation}
$$
\gm(q)\equiv \frac{\epsilon}{\pi m^{\frac{3q}{2}}} \qquad\text{and}\qquad  \Delta(q) \equiv q\pi m^q. \label{paramSW}
  \eqno{(\theequation{\mathit{a},\mathit{b}})}
$$ 
The parameter $\Delta$ is the half-width at half-maximum, and $q$ is a rational number completely determined by the restoring force. For gravity waves, $q= \frac{2}{3}$ \citep{young-wolfe}, while $q= 2$ for capillary waves. Furthermore
\be
\kd_{\star} \simeq m^{\frac{q}{2}} - \frac{q^2}{2} m^q \ln(m) + \frac{q^4}{4} m^{\frac{3q}{2}} \big[\ln(m)\big]^2- \frac{q^3}{4} m^{\frac{3q}{2}} \ln(m) , \label{kstar_exp}
\ee
which generalizes the asymptotic formula obtained by \citet{young-wolfe} using the exact solution of the Rayleigh equation. 

In the case of capillary--gravity waves, $\kcd$ and $m$ are both small parameters (see Eq. \ref{longcapgrav}), so that there exists an exponent $\nu>0$ such that $\kcd=m^{\nu}$. This exponent, originating from the choice of the control parameters, determines the value of $\kmax/\kcd$, and hence whether the fastest growing waves are driven by gravity, surface tension or both. We find that, if $\nu=\frac{1}{2}$, then $\kmax/\kcd= O(1)$, and hence the effects of gravity and surface tension play an equal role in the fastest growing waves. For $\nu=\frac{1}{2}$, we generalize the strong wind limit formula (\ref{gammaSW}) to capillary--gravity waves by taking $q=1$ and performing the transformations
\refstepcounter{equation}
$$
\gm(q) \to \frac{\gm(q)}{\xmax^2+1} \quad\text{and}\quad  \Delta(q)\to \Delta(q) Q(\xmax),
  \eqno{(\theequation{\mathit{a},\mathit{b}})} 
$$ 
where
\refstepcounter{equation}
$$
\xmax\equiv\frac{\kmax}{\kcd} \simeq \frac{119}{81}\quad\text{and}\quad Q(\xmax)\equiv  \big[\xmax^2+1\big]^{\frac{3}{2}}\frac{2 \sqrt{\xmax}}{\xmax^2+3}\ . \label{expmaxCG}
  \eqno{(\theequation{\mathit{a},\mathit{b}})} 
$$ 
We also obtain the asymptotic form of $\mu^{\rm{exp}}_{\rm{long}}$ as $m\ll1$, which is 
\be
\frac{\mu^{\rm{exp}}_{\rm{long, SW}}(\kd)}{\mu_{\rm{max}}(q)} =  \frac{2\Delta(q) [\kd-\kmax] }{\big[\kd-\kmax\big]^2+ \big[\Delta(q)\big]^2}\ , \label{muSW}
\ee
with  $\mu_{\rm{max}}(q)\equiv \frac{\gamma_{\rm{max}}}{2}$. From equations (\ref{gammaSW}) and (\ref{muSW}), we deduce that, in the strong wind limit, the graph of $\gamma$ versus $\mu$ becomes  a circle of radius $\mu_{\rm{max}}$, centered at $(0,\mu_{\rm{max}})$. 
\subsection{Logarithmic wind profile} \label{logSW}
\begin{figure*}[htbp!]
        (a)\includegraphics[trim = 0 0 0 0, clip, width = 0.52\textwidth]{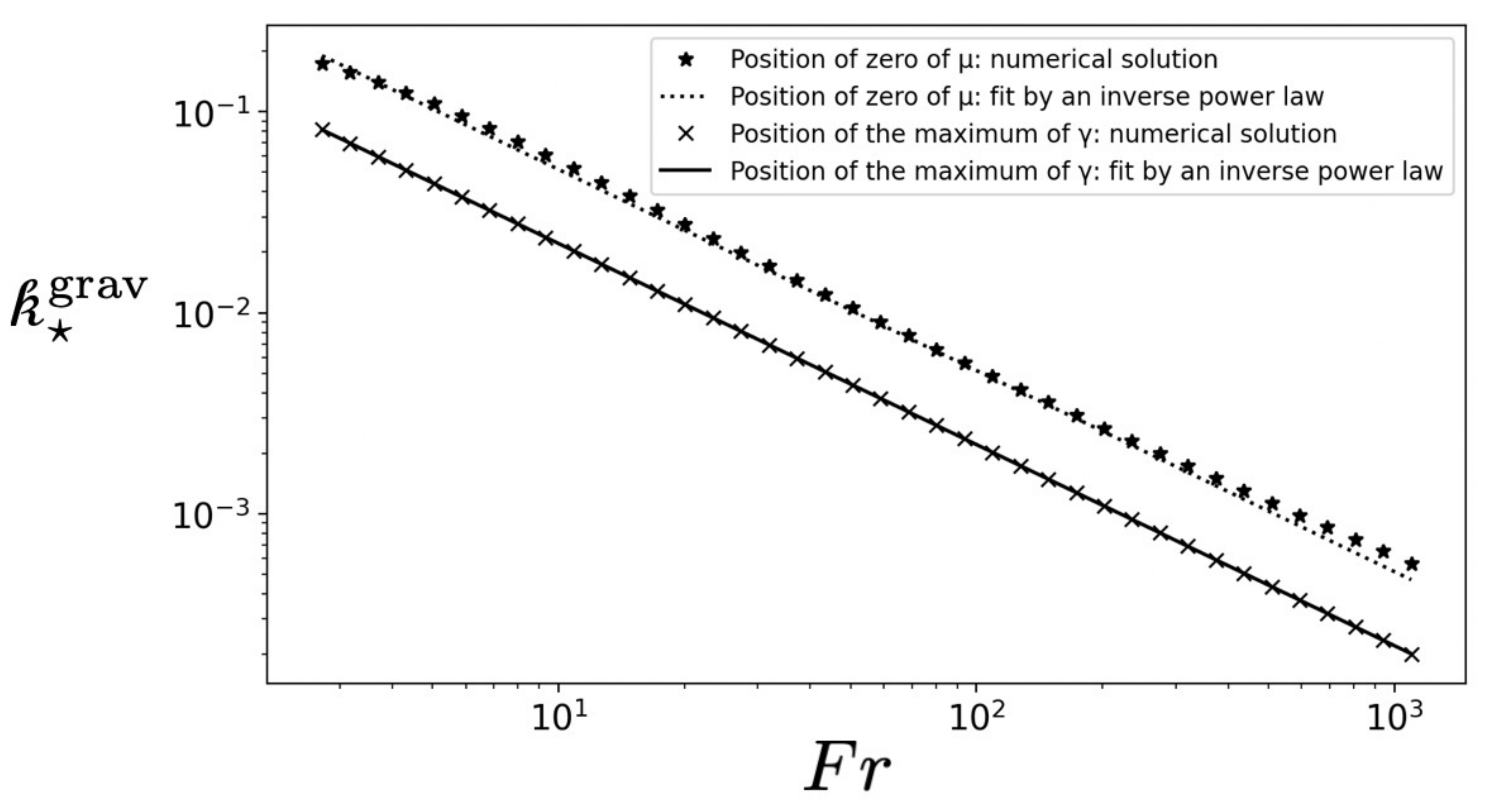}	  	
        (b)\includegraphics[trim = 0 0 0 0, clip, width = 0.38\textwidth]{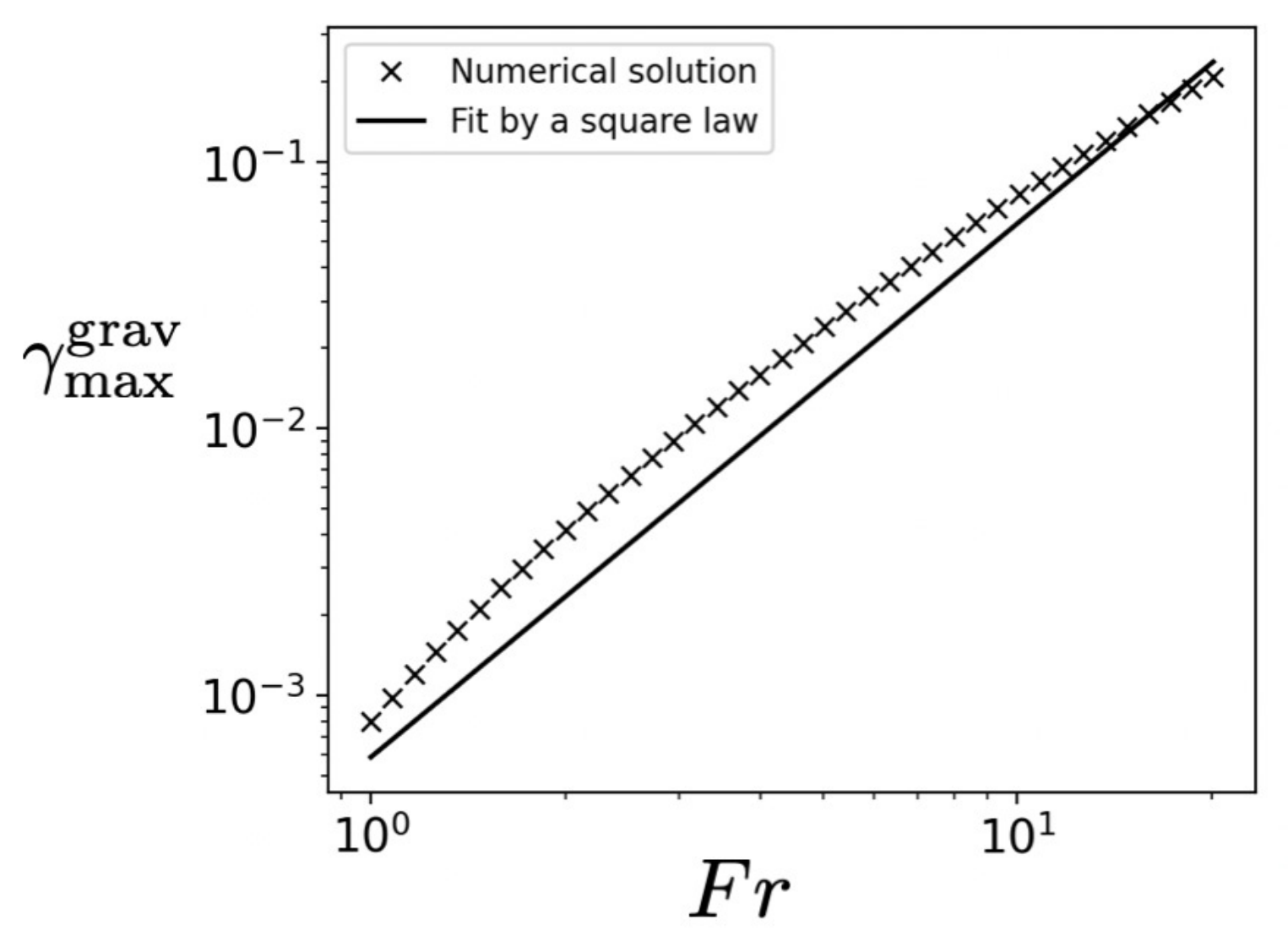}
 \caption{(a) The position of the maximum of $\gamma^{\rm{log}}_{\rm{long}}(\kd)$ and the position of the zero of $\mu^{\rm{log}}_{\rm{long}}(\kd)$, and (b) the amplitude of the maximum growth rate $ \gm^{\rm{grav}}$  for gravity waves and the logarithmic wind profile as a function of the Froude number.}
\label{powerlaws}
\end{figure*}
For $\U(\z)=\ln(1+\z)/\kappa$, we find numerically that, in the strong wind limit, the fastest growing gravity wave is determined by 
\be
\kmax^{\rm{grav}} \sim N_1\sqrt{m},\qquad m\ll1, \label{ksloggrav}
\ee
where $N_1=0.22$. Moreover, we show that the corresponding maximum growth rate is
\be
\gm^{\rm{grav}}  \sim N_2\ \frac{\epsilon}{\kappa^2\pi m}\ ,\qquad m\ll1, \label{gmloggrav}
\ee
where $N_2=0.29$. Whereas the small parameter $m= Fr^{-2}$ provides a convenient means of carrying out the asymptotic analysis, we believe that the Froude number itself provides better physical intuition. In particular, we see from Figure \ref{plantmiles} that in practice $Fr= O(10)$. Thus, we rewrite equations (\ref{ksloggrav}) and (\ref{gmloggrav}) as \refstepcounter{equation}
$$
\kmax^{\rm{grav}} \sim \frac{N_1}{Fr} \qquad\text{and}\qquad  \gm^{\rm{grav}}  \sim N_2\ \frac{\epsilon}{\kappa^2\pi}\ Fr^2. \label{powersFr}
  \eqno{(\theequation{\mathit{a},\mathit{b}})} 
$$ 
In Figure \ref{powerlaws}(a), we plot the position of the maximum of $\gamma^{\rm{log}}_{\rm{long}}(\kd)$ and the position of the zero of $\mu^{\rm{log}}_{\rm{long}}(\kd)$ as a function of $Fr$. For both quantities, the results can be fitted with an inverse power law, with the fit  for $\gamma$ slightly better than that for $\mu$. Furthermore, the two sets of points are quite close to each other, confirming that for large values of $Fr$ the position of the zero of $\mu$ is an excellent approximation to the position of the maximum of $\gamma$. Figure \ref{powerlaws}(b) shows the amplitude of the maximum growth rate as a function of $Fr$, fitted with a square law. We explain in Appendix \ref{Appmax} that (\ref{powersFr}\textit{b}) is deduced from (\ref{powersFr}\textit{a}) by assuming that $\mu=0$ determines the maximum of $\gamma$. As shown in Figure \ref{powerlaws}(a), such an assumption is very reasonable but inevitably introduces some error.
In contrast, the growth rate of capillary waves does not have a maximum, but diverges at small $\kd$ and, independent of the value of $m$,
 $\mu^{\rm{log}}_{\rm{long}}$ does not vanish. Nonetheless, the assumption that the effect of gravity is negligible does not hold for $\kd\ll\kcd$. Therefore, this divergence of $\gamma$ is not physical.

\textcolor{black}{In the case of capillary--gravity waves, as for the exponential profile, there exists an exponent $\tilde{\nu}>0$ such that $\kcd=m^{\tilde{\nu}}$. We show that $\kmax/\kcd= O(1)$ for $\tilde{\nu}=1$ (see Figure \ref{logmugamma} where $\kcd=0.8$) and 
\be
\gm^{\rm{CG}} \sim N_3\ \frac{\epsilon}{\pi \big[\kappa m\big]^2}\ ,\qquad m\ll1,\label{gmlogCG}
\ee
where $N_3$ is a numerical constant. }
Therefore, the wind-wave interaction has similar overall characteristics for both the exponential and the logarithmic wind profiles, differing only in the numerical details.

\section{Conclusion} \label{ccl}
We examine the Miles mechanism of wind-wave instability through the lens of an asymptotic analysis of the Rayleigh equation. \textcolor{black}{In the view of \citet{miles57}, free surface waves with phase speed $c_0$ perturb the wind field, and energy is transferred from the mean flow to the perturbation at the critical level, $z=z_c$, where the wind speed is equal to $c_0$. The subsequent feedback on the normal stress at the water surface leads to the growth of waves. However, the coupling with the wind field also affects the phase speed. We calculate the energy growth rate normalized by the angular frequency of free surface waves, $\gamma$, and twice the wind-dependent relative change of the phase speed, $\mu$. The emphasis is on $\mu$ and $\gamma$ being respectively proportional to the real and imaginary parts of the Fourier components of the aerodynamic pressure (see Eq. \ref{phillips}). A parameter $m$ accounts for the competition between the shear in the air and the restoring force; gravity and/or surface tension.} In the strong wind limit, defined by $m\ll1$, we find that (i) the functions $\mu=\mu(\kd)$ and $\gamma=\gamma(\kd)$ are self-similar with respect to $m$; (ii) the similarity exponents depend on the restoring force and the wind profile (see Eqs. \ref{gammaSW} and \ref{muSW} for the exponential profile); and (iii) $\gamma$ is maximal when $\mu =0$, consistent with the sheltering hypothesis of \citet{jeffreys25}. \textcolor{black}{In other words, the growth of surface waves is optimal when the aerodynamic pressure is in phase with the wave slope, and the overall instability mechanism is qualitatively independent of the strength of the wind and of the restoring force. }Additionally, we show that long waves interact with the wind only between the mean water level and the critical level, $z=\zc$. Finally, we use our asymptotic solutions to fit the entire range of data compiled by \citet{plant82}. 
\section*{Acknowledgements}
We acknowledge Swedish Research Council grant no. 638-2013-9243 for support. 
\appendix
\section{Inflexion point in the solution of the Rayleigh equation}\label{Appinflex}
\be
\kd^2+\frac{\U''(\z)}{\U(\z)-\C(\kd)}=0. \label{inflex}
\ee
Here, we solve equation (\ref{inflex}) in the small $\kd$ limit. 
\subsection{The Exponential Profile: $\U(\z) =1-e^{-\z}$}
For the exponential profile, we use the variable $\Z=\z-\zcd$ and equation (\ref{inflex}) becomes
\be
\kd^2+ \frac{1} {  1-e^{\Z}   } =0. 
\ee
Hence,
\be
\Zs = \ln\bigg(1+\frac{1}{\kd^2}\bigg)\sim-2 \ln(\kd),\qquad \kd\ll 1, \label{Zsexp}
\ee
from which we see that the matching requirement, $\kd \Zs\to 0$ as $\kd\to0$, is fulfilled. 
\subsection{The Logarithmic Profile: $\U(\z) =\ln(1+\z)/\kappa$}
For the logarithmic profile, with $\Long = \kd(1+\zcd)$ and $\Z=(1+\z)/(1+\zcd)$, equation (\ref{inflex}) takes the form
\be
\Long^2  \Z^2 \ln(\Z) =1,\qquad\Long\to0^+.\label{transc}
\ee
Equation (\ref{transc}) is transcendental, but not in a particularly useful form for perturbation theory. We let $X\equiv \Long \Z$ and divide equation (\ref{transc}) by $-\ln(\Long)$. This yields
\be
X^2\big[1+\flex \ln(X)\big] = \flex,\qquad\text{with}\qquad \flex\equiv-\frac{1}{\ln(\Long)} \to0^+. \label{tfm}
\ee
Setting $\flex=0$ on the left hand side, we obtain $X=\sqrt{ \flex}$ as a first approximation. Note that we consistently have $\flex  \ln(X)  \underset{\flex\to 0^+}{\longrightarrow} 0$. Hence, we seek a solution of the form
\be
X= \sqrt{\flex }\big[1+ a(\flex)\big],\qquad\text{with}\qquad |a(\flex)|\ll 1. \label{form}
\ee
We substitute (\ref{form}) into Eq. (\ref{tfm}) to determine $a(\flex)$,
\be
\big[1+ a(\flex)\big]^2 \Big(1+\frac{\flex}{2} \ln(\flex) + \flex \ln\big[1+ a(\flex)\big]  \Big) = 1,  \label{gen}
\ee
and since $|a(\flex)|\ll 1$, Eq. (\ref{gen}) can be simplified into 
\be
\big[1+ 2 a(\flex)\big] \Big[1+\frac{\flex}{2} \ln(\flex) + \flex a(\flex)  \Big]+ O\big[ a^2(\flex) \big] = 1.
\ee
Neglecting the higher-order term $\flex a(\flex)$, we eventually find 
\be
a(\flex) = - \frac{\flex}{4} \ln(\flex)+O\big[\flex^2\ln^2(\flex)\big],
\ee
from which we conclude that 
\be
\Zs = \frac{\ln\big|\ln(\Long)\big|-4\ln(\Long)  }{4\Long\big|\ln(\Long)\big|^{\frac{3}{2}} }, \qquad \Long\ll1.
\ee 
Finally, because the logarithm of a logarithm is an extremely slowly varying function of its argument, we discard it and arrive at the approximation
\be
\Zs \simeq \frac{1}{\Long \big|\ln(\Long)\big|^{\frac{1}{2}} }. 
\ee 
This expression captures the behaviour of $\Zs$ as $\Long$ goes to zero. In particular, it shows that $\Long\Zs$ decays very slowly to zero in the small $\Long$ limit, as needed for matching. 
\section{Approximation of $\crit$ for the exponential profile when $\zcd\ll1$}\label{Appexp}
For the exponential profile, we construct in equation (\ref{unifexp}) a uniform asymptotic approximation in terms of the variable $\Z=\z-\zcd$. Hence, the solution of the Rayleigh equation at the critical level, $\z=\zcd$, is
\be
\crit =\lim\limits_{\Z\to0}\Xu(\Z) = \kd G, \label{longcrit}
\ee
with the complex constant $G$ given in equation (\ref{Gexp}). We rewrite equation (\ref{longcrit}) as
\be
\crit^{-1}= \frac{1-e^{\zcd}}{\kd} \Big[1 - \frac{\kd}{1-e^{-\zcd}}-\kd\ln(1-e^{-\zcd})+i \kd \pi \Big].
\ee
For $z_c\ll1$, we approximate the exponential as
\be
e^{\zcd}=1+\zcd + O\big(\zcd^2\big), 
\ee
from which we readily find that
\be
\crit^{-1}\simeq 1-\frac{\zcd}{\kd}+ \zcd\ln(\zcd)-i\pi \zcd.
\ee
This expression agrees with that obtained by Miles from the exact solution in an Appendix of \citet{morland-saffman}. 
\section{Maximum growth rate in the strong wind limit}\label{Appmax}
Here, we determine the position and the amplitude of the normalized growth rate, $\gamma=\gamma(\kd)$, in the small $m$ limit (see Table \ref{tablewave}).  
\subsection{The exponential profile: $\U(\z) =1-e^{-\z}$} \label{appSWexp}
For the exponential profile, the normalized growth rate for long waves is given by equation (\ref{gammaexp}), which we write as
\be
\gamma^{\rm{exp}}_{\rm{long}}(\kd)= \frac{D(\kd)}{\big[N(\kd)\big]^2 + \big[\kd\pi\big]^2 }\ , \label{gexp}
\ee
where 
\refstepcounter{equation}
$$
D(\kd) \equiv \epsilon \pi \kd (\pp-1)^2 \quad\text{and}\quad N(\kd) \equiv 1- \kd\pp +\kd\ln(\pp). \label{DN}
  \eqno{(\theequation{\mathit{a},\mathit{b}})}
$$ 
These functions depend on the wind parameter $m$ through the dispersion relation $\pp = \pp(\kd)$. Clearly, $\gamma^{\rm{exp}}_{\rm{long}}$ becomes infinite when its denominator vanishes. Because $\kd\ll1$, we expect the maximum to occur when $N=0$. Thus for a given $\pp = \pp(\kd)$, we solve 
\be
N(\kd)=0, \qquad\text{as}\qquad m\ll1, 
\ee
and retrospectively check that the solution, $\kd_{\star}$, accurately approximates the position of the maximum of $\gamma$ as $m\ll1$. After some algebra, we obtain the expressions given in equations (\ref{kstar_exp}) and (\ref{expmaxCG}\textit{a}). Finally, we substitute the Taylor series approximations
\begin{align}
D(\kd)&=D(\kd_{\star}) + O(k-\kd_{\star}),\\
\text{and}\qquad N(\kd)& = \frac{dN}{d\kd}\bigg|_{\kmax} [\kd -\kmax] + O\big(  [\kd -\kmax] ^2\big) \label{taylor}
\end{align}
into equation (\ref{gexp}), and obtain the Lorentzian function given in equation (\ref{gammaSW}).

Noting that 
\be
\mu^{\rm{exp}}_{\rm{long}} (\kd)= -\epsilon (\pp-1)^2 \frac{N(\kd)}{\big[N(\kd)\big]^2 + \big[\kd\pi\big]^2 }\ ,
\ee
we use equation (\ref{taylor}) to determine the asymptotic form of $\mu^{\rm{exp}}_{\rm{long}}$ given in equation (\ref{muSW}). 
\subsection{The logarithmic profile: $\U(\z) =\ln(1+\z)/\kappa$}
For the logarithmic profile, the normalized growth rate for long waves is given by equation (\ref{gammalog}), where the dependence on the wind parameter $m$ comes from the critical level, $\zcd$. We proceed as in \S \ref{appSWexp}: for a given function $\zcd=\zcd(\kd)$, we numerically solve
\be
H(\Zs)g(\zcd) -J(\Zs)f(\zcd)=0. \label{Hg-Jf}
\ee
There is no solution in the case of capillary waves. For gravity waves and $m\ll1$, we extract the power law given in equation (\ref{ksloggrav}), and use it to approximate $\gamma^{\rm{log}}_{\rm{long}} (\kd_{\star})$; with the aid of equation (\ref{Hg-Jf}), we eventually obtain equation (\ref{gmloggrav}). The procedure is similar for capillary--gravity waves. 
\section{Squire's theorem for wind waves}\label{Appsquire}
For an inviscid flow between two fixed boundaries, Squire's theorem states that, to each three-dimensional disturbance, there corresponds a more unstable two-dimensional one \citep{drazin-reid}. \citet{morland-saffman} suggest an extension of the Squire transformation to gravity waves propagating in a direction different from the one in which the wind blows. They conclude that the theorem holds for the exponential profile but fails for the logarithmic profile. Here, we revisit their work and include the effect of surface tension.

Let $\bb{\hat{k}}$ be the unit vector defining the direction of wave propagation and orient $\bb{\hat{x}}$ in the direction of the mean wind field.  
We rotate the coordinate system by the angle $\theta\equiv (\bb{\hat{x}},\bb{\hat{k}})$ \citep{linbook}
and show that only the component of the base-state flow in the direction of $\bb{\hat{k}}$ affects the disturbance. Therefore, we need only perform the transformation
\be
V\to V\cos(\theta),
\ee
where $V$ is the velocity scale associated with the wind profile, which gives 
\be
 m \to \frac{m}{\cos^2(\theta)}\ . \label{squireTF}
\ee
For the exponential profile, we see from equation (\ref{paramSW}) that the transformation (\ref{squireTF}) reduces the maximum growth rate, regardless of the restoring force. In the case of the logarithmic profile, the growth rate only has a maximum for gravity and capillary--gravity waves. Then, from equations (\ref{gmloggrav}) and (\ref{gmlogCG}), we also infer a reduction in the maximum under the transformation (\ref{squireTF}). 

We conclude that for both wind profiles three-dimensional perturbations have a maximum growth rate of smaller amplitude than their two-dimensional counterparts. Hence, Squire's theorem extends to wind waves. The reason why \citet{morland-saffman} found that it does not hold in the case of the logarithmic profile is that they used the viscous length scale instead of the roughness length as the characteristic length scale $L$. 
 \section{Fixed points of the air flow}\label{Appfixedpoint}
A fixed point of the flow is given by $\bb{u}(x,z,t)= \bb{0}$. After the expansions (\ref{expansions}\textit{a,b}), the streamfunction of the air flow at the leading order in $\epsilon$ is 
\be
\psi^{\rm{a}}_0(x,z,t) =\Re\big\{ \hat{\psi}_0(z)\ e^{ik(x-c_0 t)}\big\}.
\ee
Because of the singularity at $z=\zc$, $\hat{\psi}_0$ is a complex function. We denote $\psir $ and $\psii$ its real and imaginary parts, respectively, and let $\x\equiv x-c_0 t$ be the horizontal coordinate in a frame moving at speed $c_0$. Then, the components of the velocity field are 
\begin{align}
u(\x,z)&=\psir'(z) \cos(k\x) - \psii'(z) \sin(k\x),\\
\text{and}\quad w(\x,z)&=k\psir(z) \sin(k\x) + k\psii(z)\cos(k\x),
\end{align}
where again the prime denotes differentiation with respect to $z$. Here, we seek points $(\xe,\ze)$ such that
\be
\bc
u(\xe,\ze)=0\\
w(\xe,\ze)=0
\ec,
\ee
and study their stability. Here, we have calculated the function $\chi(\z)= \hat{\psi}_0(\z)/\hat{\psi}_0(0)$, where $\z$ is the dimensionless version of $z$. We now show that there is a unique point $\zex$ located between $\zcd$ and $\zs$ (the position of the inflexion point) where both the real and imaginary parts of $\chi$ have an extremum, which implies $u(\x,\ze)=0$ for any $\x$. We determine the nature of this extremum for a given value of $k$. Additionally, we prove that the fixed points are elliptic. 
\subsection{The Exponential Profile: $\U(\z) =1-e^{-\z}$} \label{derexp}
For the exponential profile, we make the ansatz that the uniform asymptotic approximation $\Xu(\Z)$ has an extremum at $\Ze\gg1$, which we check {\em a-posteriori.} The derivatives for $\Z\gg 1$ and $\kd\ll1$ are 
\begin{align}
\Xu'(\Z)&\sim G\ e^{-\Z}\big[1-\kd(e^{\Z}+\Z)\big],\\
\text{and}\quad \Xu''(\Z)&\sim-G\ e^{-\Z}\big[\kd e^{\Z}+1-\kd(e^{\Z}+\Z)\big]. \label{D2exp}
\end{align}
Hence, $\Xu$ has an extremum at the point $\Ze$, the solution of
\be
e^{\Z} + \Z =\frac{1}{\kd}, \qquad \kd\to0^+. \label{extexp}
\ee
Using a similar method to that given in Appendix \ref{Appinflex}, we show that
\be
\Ze=\ln\bigg\{\frac{1}{\kd}+\ln(\kd) + O\big[\kd\ln(\kd)\big]\bigg\}. 
\ee
The position  of the inflexion point, $\Zs$, is given in equation (\ref{Zsexp}), thus
\be
1\ll\Ze\ll \Zs, \qquad\text{as}\qquad \kd\ll1. 
\ee
Using equations (\ref{D2exp}) and (\ref{extexp}), together with the expression for $G$ (see Eq. \ref{Gexp}), we obtain
\be
\Xu''(\Zs)\sim \kd(\pp-1) \frac{1-\kd \pp + \kd\ln(\pp)-i\kd\pi}{[1-\kd \pp + \kd\ln(\pp)]^2 + [\kd \pi]^2}\ .
\ee
Because $\pp>1$, we conclude that
\begin{align}
\text{Sign}\Big(&\Re\big\{\Xu''(\Zs) \big\}\Big)= - \text{Sign} \Big(\mu^{\rm{exp}}_{\rm{long}}\Big),\\
\text{and}\qquad &\Im\big\{\Xu''(\Zs) \big\}<0.
\end{align}
\subsection{The Logarithmic Profile: $\U(\z) =\ln(1+\z)/\kappa$} \label{derlog}
For the logarithmic profile, the derivatives of the outer solution for $\Z>1$ are 
\begin{align}
\Xout'(\Z)&=\frac{J-H \li(\Z)}{\Zs\Z}\ G\ e^{-\Long \Zs},\\
\text{and}\quad \Xout''(\Z)&=-\frac{H\Z+\ln(\Z)[J-H \li(\Z)]}{\Zs\Z^2\ln(\Z)}\ G\ e^{-\Long \Zs}.\label{D2log}
\end{align}
Hence, $\Xout$ has an extremum at a point $\Ze$ such that 
\be
\li(\Ze) = \li(\Zs)-\frac{\Long \Zs^2}{H}\ . \label{extlog}
\ee
Because the logarithmic integral function is monotonic, we have $\Ze<\Zs$. Using equations (\ref{D2log}) and (\ref{extlog}) together with the expression for $G$ (see Eq. \ref{Glog}), we obtain
\be
\Xout''(\Zs) = - \frac{H (1+\zcd)}{\Ze\ln(\Ze)}\ \frac{ Hg-Jf + i\pi HF}{\big[ Hg-Jf\big]^2 +\big[\pi Hf \big]^2}\ ,
\ee
from which we conclude that
\begin{align}
\text{Sign}\Big(&\Re\big\{\Xout''(\Zs) \big\}\Big)= - \text{Sign} \Big(\mu^{\rm{exp}}_{\rm{long}}\Big),\\
\text{and}\qquad &\Im\big\{\Xout''(\Zs) \big\}<0.
\end {align}
\subsection{The Fixed Points are Elliptic}
Having calculated $\ze$ and now need to determine $\xe$ such that $w(\xe,\ze)=0$. Using the polar form of $\hat{\psi}_0(z)=|\hat{\psi}_0(z)| e^{i\varphi(z)}$, we have 
\be
w(\x,z) = k |\hat{\psi}_0(z)| \sin[k\x+\varphi(z)]. 
\ee
As a consequence, for a given $k$ there are two possible values for $\xe$, defined as follows:
 \refstepcounter{equation}
$$
k\xeo +\varphi(\ze)=\pi \quad\text{and}\quad k\xet+\varphi(\ze)=0. \label{xe}
\eqno{(\theequation{\mathit{a},\mathit{b}})}
$$ 
For both wind profiles, the phase of $\chi(\ze)$ is the same as the phase of $G$. Therefore, $\varphi(\ze)=\pi/2$ when $k$ is equal to $k_{\star}$, the wavenumber of the fastest growing wave. Moreover
\be
\varphi(\ze)>\frac{\pi}{2}\quad\text{for}\quad k<k_{\star},\quad\text{and}\quad\varphi(\ze)<\frac{\pi}{2}\quad\text{otherwise.} \label{zephase}
\ee
In order to determine the stability of the fixed point we have just found, we study the eigenvalues of the gradient velocity matrix. Using 
\be
\cos[k\xej+\varphi(\ze)]=(-1)^j, \qquad j=1,2,
\ee
we obtain
\begin{widetext}
\be
\bb{\nabla}\bb{u}(\xej, \ze)=
\begin{pmatrix}
0&\psir''(\ze)\cos(k\xej)-\psii''(\ze)\sin(k\xej) \\ k^2|\psi_0(\ze)| (-1)^j &0
\end{pmatrix},\qquad j = 1,2. 
\ee
\end{widetext}
The eigenvalues are the roots of the characteristic polynomial, 
\be
X^2 + \det\bb{\nabla}\bb{u}(\xej, \ze)=0.
\ee
Using equations (\ref{xe}) and (\ref{zephase}), together with the results of \S \ref{derexp} and \S \ref{derlog}, we check that in all cases 
\be
\det\bb{\nabla}\bb{u}(\xej, \ze)>0.
\ee
Thus, the eigenvalues are purely imaginary, and complex conjugates of each other. Therefore, the fixed points are elliptic.

%


\begin{thebibliography}{50}%
\makeatletter
\providecommand \@ifxundefined [1]{%
 \@ifx{#1\undefined}
}%
\providecommand \@ifnum [1]{%
 \ifnum #1\expandafter \@firstoftwo
 \else \expandafter \@secondoftwo
 \fi
}%
\providecommand \@ifx [1]{%
 \ifx #1\expandafter \@firstoftwo
 \else \expandafter \@secondoftwo
 \fi
}%
\providecommand \natexlab [1]{#1}%
\providecommand \enquote  [1]{``#1''}%
\providecommand \bibnamefont  [1]{#1}%
\providecommand \bibfnamefont [1]{#1}%
\providecommand \citenamefont [1]{#1}%
\providecommand \href@noop [0]{\@secondoftwo}%
\providecommand \href [0]{\begingroup \@sanitize@url \@href}%
\providecommand \@href[1]{\@@startlink{#1}\@@href}%
\providecommand \@@href[1]{\endgroup#1\@@endlink}%
\providecommand \@sanitize@url [0]{\catcode `\\12\catcode `\$12\catcode
  `\&12\catcode `\#12\catcode `\^12\catcode `\_12\catcode `\%12\relax}%
\providecommand \@@startlink[1]{}%
\providecommand \@@endlink[0]{}%
\providecommand \url  [0]{\begingroup\@sanitize@url \@url }%
\providecommand \@url [1]{\endgroup\@href {#1}{\urlprefix }}%
\providecommand \urlprefix  [0]{URL }%
\providecommand \Eprint [0]{\href }%
\providecommand \doibase [0]{http://dx.doi.org/}%
\providecommand \selectlanguage [0]{\@gobble}%
\providecommand \bibinfo  [0]{\@secondoftwo}%
\providecommand \bibfield  [0]{\@secondoftwo}%
\providecommand \translation [1]{[#1]}%
\providecommand \BibitemOpen [0]{}%
\providecommand \bibitemStop [0]{}%
\providecommand \bibitemNoStop [0]{.\EOS\space}%
\providecommand \EOS [0]{\spacefactor3000\relax}%
\providecommand \BibitemShut  [1]{\csname bibitem#1\endcsname}%
\let\auto@bib@innerbib\@empty

\bibitem [{\citenamefont {Miles}(1957)}]{miles57}%
  \BibitemOpen
  \bibfield  {author} {\bibinfo {author} {\bibfnamefont {J.~W.}\ \bibnamefont
  {Miles}},\ }\href@noop {} {\bibfield  {journal} {\bibinfo  {journal}
  {J.~Fluid Mech.}\ }\textbf {\bibinfo {volume} {3}},\ \bibinfo {pages} {185}
  (\bibinfo {year} {1957})}\BibitemShut {NoStop}%
\bibitem [{\citenamefont {von Helmholtz}(1868)}]{helmholtz1868}%
  \BibitemOpen
  \bibfield  {author} {\bibinfo {author} {\bibfnamefont {H.}~\bibnamefont {von
  Helmholtz}},\ }\href@noop {} {\bibfield  {journal} {\bibinfo  {journal}
  {Monatsberichte der K{\"o}niglichen Preussische Akademie der Wissenschaften
  zu Berlin}\ }\textbf {\bibinfo {volume} {23}},\ \bibinfo {pages} {215}
  (\bibinfo {year} {1868})}\BibitemShut {NoStop}%
\bibitem [{\citenamefont {Thomson}(1871)}]{kelvin1871}%
  \BibitemOpen
  \bibfield  {author} {\bibinfo {author} {\bibfnamefont {W.~L.~K.}\
  \bibnamefont {Thomson}},\ }\href@noop {} {\bibfield  {journal} {\bibinfo
  {journal} {Philosophical Magazine}\ }\textbf {\bibinfo {volume} {42}},\
  \bibinfo {pages} {362} (\bibinfo {year} {1871})}\BibitemShut {NoStop}%
\bibitem [{\citenamefont {Jeffreys}(1925)}]{jeffreys25}%
  \BibitemOpen
  \bibfield  {author} {\bibinfo {author} {\bibfnamefont {H.}~\bibnamefont
  {Jeffreys}},\ }\href@noop {} {\bibfield  {journal} {\bibinfo  {journal}
  {Proc. R. Soc. Lond. A}\ }\textbf {\bibinfo {volume} {107}},\ \bibinfo
  {pages} {189} (\bibinfo {year} {1925})}\BibitemShut {NoStop}%
\bibitem [{\citenamefont {Phillips}(1957)}]{phillips57}%
  \BibitemOpen
  \bibfield  {author} {\bibinfo {author} {\bibfnamefont {O.~M.}\ \bibnamefont
  {Phillips}},\ }\href@noop {} {\bibfield  {journal} {\bibinfo  {journal}
  {J.~Fluid Mech.}\ }\textbf {\bibinfo {volume} {2}},\ \bibinfo {pages} {417}
  (\bibinfo {year} {1957})}\BibitemShut {NoStop}%
\bibitem [{\citenamefont {Phillips}(1977)}]{phillipsbook}%
  \BibitemOpen
  \bibfield  {author} {\bibinfo {author} {\bibfnamefont {O.~M.}\ \bibnamefont
  {Phillips}},\ }\href@noop {} {\emph {\bibinfo {title} {The Dynamics of the
  Upper Ocean}}}\ (\bibinfo  {publisher} {Cambridge University Press},\
  \bibinfo {year} {1977})\BibitemShut {NoStop}%
\bibitem [{\citenamefont {Janssen}(2004)}]{janssenbook}%
  \BibitemOpen
  \bibfield  {author} {\bibinfo {author} {\bibfnamefont {P.~A. E.~M.}\
  \bibnamefont {Janssen}},\ }\href@noop {} {\emph {\bibinfo {title} {The
  Interaction of Ocean Waves and Wind}}}\ (\bibinfo  {publisher} {Cambridge
  University Press},\ \bibinfo {address} {Cambridge, UK},\ \bibinfo {year}
  {2004})\BibitemShut {NoStop}%
\bibitem [{\citenamefont {Drazin}\ and\ \citenamefont
  {Reid}(1981)}]{drazin-reid}%
  \BibitemOpen
  \bibfield  {author} {\bibinfo {author} {\bibfnamefont {P.~G.}\ \bibnamefont
  {Drazin}}\ and\ \bibinfo {author} {\bibfnamefont {W.~H.}\ \bibnamefont
  {Reid}},\ }\href@noop {} {\emph {\bibinfo {title} {Hydrodynamic stability}}}\
  (\bibinfo  {publisher} {Cambridge University Press},\ \bibinfo {year}
  {1981})\BibitemShut {NoStop}%
\bibitem [{\citenamefont {Conte}\ and\ \citenamefont
  {Miles}(1959)}]{conte-miles}%
  \BibitemOpen
  \bibfield  {author} {\bibinfo {author} {\bibfnamefont {S.~D.}\ \bibnamefont
  {Conte}}\ and\ \bibinfo {author} {\bibfnamefont {J.~W.}\ \bibnamefont
  {Miles}},\ }\href@noop {} {\bibfield  {journal} {\bibinfo  {journal} {J. Soc.
  Indust. App. Math.}\ }\textbf {\bibinfo {volume} {7}},\ \bibinfo {pages}
  {361} (\bibinfo {year} {1959})}\BibitemShut {NoStop}%
\bibitem [{\citenamefont {Hughes}\ and\ \citenamefont
  {Reid}(1965)}]{hughes-reid}%
  \BibitemOpen
  \bibfield  {author} {\bibinfo {author} {\bibfnamefont {T.~H.}\ \bibnamefont
  {Hughes}}\ and\ \bibinfo {author} {\bibfnamefont {W.~H.}\ \bibnamefont
  {Reid}},\ }\href@noop {} {\bibfield  {journal} {\bibinfo  {journal} {J.~Fluid
  Mech.}\ }\textbf {\bibinfo {volume} {23}},\ \bibinfo {pages} {717} (\bibinfo
  {year} {1965})}\BibitemShut {NoStop}%
\bibitem [{\citenamefont {Beji}\ and\ \citenamefont
  {Nadaoka}(2004)}]{beji-nadaoka}%
  \BibitemOpen
  \bibfield  {author} {\bibinfo {author} {\bibfnamefont {S.}~\bibnamefont
  {Beji}}\ and\ \bibinfo {author} {\bibfnamefont {K.}~\bibnamefont {Nadaoka}},\
  }\href@noop {} {\bibfield  {journal} {\bibinfo  {journal} {J.~Fluid Mech.}\
  }\textbf {\bibinfo {volume} {500}},\ \bibinfo {pages} {65} (\bibinfo {year}
  {2004})}\BibitemShut {NoStop}%
\bibitem [{\citenamefont {Young}\ and\ \citenamefont
  {Wolfe}(2013)}]{young-wolfe}%
  \BibitemOpen
  \bibfield  {author} {\bibinfo {author} {\bibfnamefont {W.~R.}\ \bibnamefont
  {Young}}\ and\ \bibinfo {author} {\bibfnamefont {C.~L.}\ \bibnamefont
  {Wolfe}},\ }\href@noop {} {\bibfield  {journal} {\bibinfo  {journal}
  {J.~Fluid Mech.}\ }\textbf {\bibinfo {volume} {739}},\ \bibinfo {pages} {276}
  (\bibinfo {year} {2013})}\BibitemShut {NoStop}%
\bibitem [{\citenamefont {Miles}(1993)}]{miles93}%
  \BibitemOpen
  \bibfield  {author} {\bibinfo {author} {\bibfnamefont {J.~W.}\ \bibnamefont
  {Miles}},\ }\href@noop {} {\bibfield  {journal} {\bibinfo  {journal}
  {J.~Fluid Mech.}\ }\textbf {\bibinfo {volume} {256}},\ \bibinfo {pages} {427}
  (\bibinfo {year} {1993})}\BibitemShut {NoStop}%
\bibitem [{\citenamefont {Plant}(1982)}]{plant82}%
  \BibitemOpen
  \bibfield  {author} {\bibinfo {author} {\bibfnamefont {W.~J.}\ \bibnamefont
  {Plant}},\ }\href@noop {} {\bibfield  {journal} {\bibinfo  {journal} {J.
  Geophys. Res.}\ }\textbf {\bibinfo {volume} {87}},\ \bibinfo {pages} {1961}
  (\bibinfo {year} {1982})}\BibitemShut {NoStop}%
\bibitem [{\citenamefont {Miles}(1959)}]{milespart2}%
  \BibitemOpen
  \bibfield  {author} {\bibinfo {author} {\bibfnamefont {J.~W.}\ \bibnamefont
  {Miles}},\ }\href@noop {} {\bibfield  {journal} {\bibinfo  {journal}
  {J.~Fluid Mech.}\ }\textbf {\bibinfo {volume} {6}},\ \bibinfo {pages} {568}
  (\bibinfo {year} {1959})}\BibitemShut {NoStop}%
\bibitem [{\citenamefont {Lighthill}(1962)}]{lighthill62}%
  \BibitemOpen
  \bibfield  {author} {\bibinfo {author} {\bibfnamefont {M.~J.~W.}\
  \bibnamefont {Lighthill}},\ }\href@noop {} {\bibfield  {journal} {\bibinfo
  {journal} {J.~Fluid Mech.}\ }\textbf {\bibinfo {volume} {14}},\ \bibinfo
  {pages} {385} (\bibinfo {year} {1962})}\BibitemShut {NoStop}%
\bibitem [{\citenamefont {Hristov}\ \emph {et~al.}(2003)\citenamefont
  {Hristov}, \citenamefont {Miller},\ and\ \citenamefont
  {Friehe}}]{hristov-nature}%
  \BibitemOpen
  \bibfield  {author} {\bibinfo {author} {\bibfnamefont {T.}~\bibnamefont
  {Hristov}}, \bibinfo {author} {\bibfnamefont {S.}~\bibnamefont {Miller}}, \
  and\ \bibinfo {author} {\bibfnamefont {C.}~\bibnamefont {Friehe}},\
  }\href@noop {} {\bibfield  {journal} {\bibinfo  {journal} {Nature}\ }\textbf
  {\bibinfo {volume} {422}},\ \bibinfo {pages} {55-58} (\bibinfo {year} {2003})}\BibitemShut {NoStop}%
\bibitem [{\citenamefont {Carpenter}\ \emph {et~al.}(2017)\citenamefont
  {Carpenter}, \citenamefont {Gua},\ and\ \citenamefont
  {Heifetz}}]{carpenter-et-al17}%
  \BibitemOpen
  \bibfield  {author} {\bibinfo {author} {\bibfnamefont {J.~R.}\ \bibnamefont
  {Carpenter}}, \bibinfo {author} {\bibfnamefont {A.}~\bibnamefont {Gua}}, \
  and\ \bibinfo {author} {\bibfnamefont {E.}~\bibnamefont {Heifetz}},\
  }\href@noop {} {\bibfield  {journal} {\bibinfo  {journal} {J. Phys.
  Oceanog.}\ }\textbf {\bibinfo {volume} {47}},\ \bibinfo {pages} {1441}
  (\bibinfo {year} {2017})}\BibitemShut {NoStop}%
\bibitem [{\citenamefont {Morland}\ and\ \citenamefont
  {Saffman}(1992)}]{morland-saffman}%
  \BibitemOpen
  \bibfield  {author} {\bibinfo {author} {\bibfnamefont {L.~C.}\ \bibnamefont
  {Morland}}\ and\ \bibinfo {author} {\bibfnamefont {P.~G.}\ \bibnamefont
  {Saffman}},\ }\href@noop {} {\bibfield  {journal} {\bibinfo  {journal}
  {J.~Fluid Mech.}\ }\textbf {\bibinfo {volume} {252}},\ \bibinfo {pages} {383}
  (\bibinfo {year} {1992})}\BibitemShut {NoStop}%
\bibitem [{\citenamefont {Wu}(1975)}]{wu75}%
  \BibitemOpen
  \bibfield  {author} {\bibinfo {author} {\bibfnamefont {J.}~\bibnamefont
  {Wu}},\ }\href@noop {} {\bibfield  {journal} {\bibinfo  {journal} {J.~Fluid
  Mech.}\ }\textbf {\bibinfo {volume} {68}},\ \bibinfo {pages} {49} (\bibinfo
  {year} {1975})}\BibitemShut {NoStop}%
\bibitem [{\citenamefont {Lin}(1955)}]{linbook}%
  \BibitemOpen
  \bibfield  {author} {\bibinfo {author} {\bibfnamefont {C.~C.}\ \bibnamefont
  {Lin}},\ }\href@noop {} {\emph {\bibinfo {title} {The Theory of Hydrodynamic
  Stability}}}\ (\bibinfo  {publisher} {Cambridge University Press},\ \bibinfo
  {year} {1955})\BibitemShut {NoStop}%
\bibitem [{\citenamefont {Bender}\ and\ \citenamefont {Orszag}(1999)}]{B-O}%
  \BibitemOpen
  \bibfield  {author} {\bibinfo {author} {\bibfnamefont {C.~M.}\ \bibnamefont
  {Bender}}\ and\ \bibinfo {author} {\bibfnamefont {S.~A.}\ \bibnamefont
  {Orszag}},\ }\href@noop {} {\emph {\bibinfo {title} {Advanced Mathematical
  Methods for Scientists and Engineers}}}\ (\bibinfo  {publisher} {Springer},\
  \bibinfo {year} {1999})\BibitemShut {NoStop}%
\bibitem [{\citenamefont {Larson}\ and\ \citenamefont
  {Wright}(1975)}]{larson-wright}%
  \BibitemOpen
  \bibfield  {author} {\bibinfo {author} {\bibfnamefont {T.~R.}\ \bibnamefont
  {Larson}}\ and\ \bibinfo {author} {\bibfnamefont {J.~W.}\ \bibnamefont
  {Wright}},\ }\href@noop {} {\bibfield  {journal} {\bibinfo  {journal}
  {J.~Fluid Mech.}\ }\textbf {\bibinfo {volume} {70}},\ \bibinfo {pages} {417}
  (\bibinfo {year} {1975})}\BibitemShut {NoStop}%
\bibitem [{\citenamefont {Banner}\ and\ \citenamefont
  {Melville}(1976)}]{banner-melville}%
  \BibitemOpen
  \bibfield  {author} {\bibinfo {author} {\bibfnamefont {M.~L.}\ \bibnamefont
  {Banner}}\ and\ \bibinfo {author} {\bibfnamefont {W.~K.}\ \bibnamefont
  {Melville}},\ }\href@noop {} {\bibfield  {journal} {\bibinfo  {journal}
  {J.~Fluid Mech.}\ }\textbf {\bibinfo {volume} {77}},\ \bibinfo {pages} {825}
  (\bibinfo {year} {1976})}\BibitemShut {NoStop}%
\bibitem [{\citenamefont {Stewart}(1974)}]{stewart74}%
  \BibitemOpen
  \bibfield  {author} {\bibinfo {author} {\bibfnamefont {R.~W.}\ \bibnamefont
  {Stewart}},\ }\href@noop {} {\bibfield  {journal} {\bibinfo  {journal}
  {Boundary-Layer Meteorology}\ }\textbf {\bibinfo {volume} {6}},\ \bibinfo
  {pages} {151} (\bibinfo {year} {1974})}\BibitemShut {NoStop}%

\end{thebibliography}

%

\end{document}